\documentclass[article]{JHEP3}

\usepackage{amssymb}
\usepackage{amsmath}
\usepackage{amsfonts}
\usepackage{graphicx}
\usepackage{epsfig}

\providecommand{\U}[1]{\protect\rule{.1in}{.1in}}

\def\spaup#1{\phantom{\fbox{\rule[#1cm]{0cm}{0cm}}}}
\def\spadown#1{\phantom{\fbox{\rule[-#1cm]{0cm}{0cm}}}}

\newcommand{\be}{\begin{equation}}
\newcommand{\ee}{\end{equation}}
\newcommand{\bea}{\begin{eqnarray}}
\newcommand{\eea}{\end{eqnarray}}
\newcommand{\bean}{\begin{eqnarray*}}
\newcommand{\eean}{\end{eqnarray*}}

\def\beq{\begin{equation}}

\def\eeq{\end{equation}}

\relax

\preprint{
{\small{\textsf{ROM2F/2006/26}}}
\\
{\small{\textsf{LPTENS-06/50}}}
\\
{\small{\textsf{CERN-PH-TH/2006-234}}}
}

\title{Eikonal Approximation in AdS/CFT:\\
Conformal Partial Waves and Finite N  Four--Point Functions}

\author{Lorenzo Cornalba$^a$, Miguel S. Costa$^{b,c}$, Jo\~ao Penedones$^{b,c}$ and Ricardo Schiappa$^d$
\\
\\
$^{a}$Dipartimento di Fisica \& INFN, Universit\'{a} di Roma ``Tor Vergata'',\\
Via della Ricerca Scientifica 1, 00133, Roma, Italy\\
\\
$^{b}$Departamento de F\'{i}sica e Centro de F\'{i}sica do Porto,\\
Faculdade de Ci\^{e}ncias da Universidade do Porto,\\
Rua do Campo Alegre, 687, 4169--007 Porto, Portugal\\
\\
$^{c}$Laboratoire de Physique Th\'eorique de l'Ecole Normale Sup\'erieure\footnote{Unit\'e mixte du C.N.R.S. et de l'Ecole Normale Sup\'erieure, UMR 8549.},\\
24 Rue Lhomond, 75231 Paris, France\\
\\
$^{d}$Theory Division, Department of Physics, CERN,\\
CH--1211 Gen\`eve 23, Switzerland\\
\\
\email{cornalba@roma2.infn.it}, \quad
\email{miguelc@fc.up.pt}, \quad
\email{jpenedones@fc.up.pt}, \quad
\email{ricardos@mail.cern.ch}
}

\abstract{We introduce the impact parameter representation for
conformal field theory correlators of the form $\mathcal{A}\sim
\left\langle\mathcal{O}_{1}\mathcal{O}_{2}\mathcal{O}_{1}\mathcal{O}_{2}\right\rangle$.
This representation is appropriate in the eikonal kinematical
regime, and approximates the conformal partial wave decomposition
in the limit of large spin and dimension of the exchanged primary.
Using recent results on the two--point function $\left\langle
\mathcal{O}_{1}\mathcal{O}_{1}\right\rangle_{\textrm{shock}}$ in
the presence of a shock wave in Anti--de Sitter, and its relation
to the discontinuity of the four--point amplitude $\mathcal{A}$
across a kinematical branch cut, we find  the high spin and
dimension conformal partial wave decomposition of all tree--level
Anti--de Sitter Witten diagrams. We show that, as in flat space,
the eikonal kinematical regime is dominated by the
\emph{T}--channel exchange of the massless particle with highest
spin ({\em graviton dominance}). We also compute the anomalous
dimensions of the high spin $\mathcal{O}_{1}\mathcal{O}_{2}$
composites. Finally, we conjecture a formula re--summing
crossed--ladder Witten diagrams to \textit{all} orders in the
gravitational coupling.}

\begin{document}



\vfill

\eject


\section{Introduction and Summary}


The AdS/CFT correspondence opens the fascinating possibility of
describing
quantum gravitational interactions using gauge theories and vice--versa \cite%
{Malda1, WittenGubser, w98,fmmr98, Malda2}. When the radius $\ell
$ of Anti--de Sitter (AdS) is much larger than the string length,
the correspondence relates a local gravitational theory,
describing the dynamics of string massless modes in AdS, to a
strongly coupled gauge theory defined on the AdS boundary. The
perturbative expansion in the AdS gravitational coupling $G$
corresponds, in general, to the $1/N$ expansion in the gauge
theory. However, the understanding of loop corrections in AdS
quantum gravity is presently beyond our reach and it is therefore
difficult to check the correspondence beyond the planar limit or
to use it to derive finite $N$ effects of strongly coupled gauge
theories. In flat space, where similar problems are present, it is
nonetheless possible to re--sum the gravitational loop expansion
when considering four--point amplitudes in the specific eikonal
kinematical regime. In a companion paper \cite{ourSW}, we
generalize the eikonal approximation to the calculation of
four--point amplitudes in AdS spaces. In the sequel, we shall
explore the consequences of these results for the dual conformal
field theory (CFT), therefore initiating a program to probe
\textit{finite} $N$ effects of strongly coupled gauge theories.

We shall concentrate on CFT correlators like $\mathcal{A}\sim
\left\langle
\mathcal{O}_{1}\mathcal{O}_{2}\mathcal{O}_{1}\mathcal{O}_{2}\right\rangle
$, which can in general be decomposed in conformal partial waves
in various channels, thus giving information about the CFT
spectrum and three--point couplings. Neglecting string
corrections, these correlators can be computed gravitationally
using Witten diagrams, which describe interactions in AdS.
Unfortunately, it is a hard problem to decompose a given AdS
diagram in conformal partial waves and therefore to improve our
understanding of the dual CFT. Recall that, in flat space, the
eikonal kinematical regime of small scattering angle controls the
large spin behavior of the partial wave decomposition.
Analogously, we shall use the results derived in \cite{ourSW} to
prove a similar result in AdS. We will show, in particular, how to
determine directly the high spin and dimension decomposition of
tree--level Witten diagrams and how to compute the anomalous
dimensions of certain high spin composite operators.

\begin{figure}
[ptb]
\begin{center}
\includegraphics[height=1.2503in,width=3.1685in]
{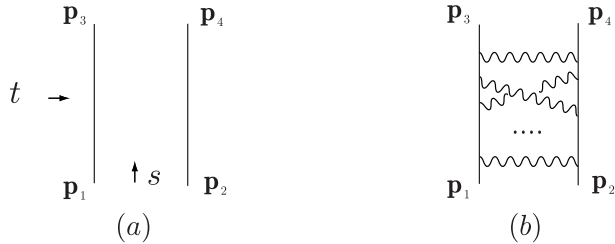}
\caption{In the eikonal regime, free propagation $\left( a\right)$ is
modified primarily by interactions described by crossed--ladder graphs $\left( b\right)$.}
\label{fig4}
\end{center}
\end{figure}

Let us recall some relevant facts on the eikonal formalism in flat
space.
Consider the scattering of two scalar particles in flat Minkowski space $%
\mathbb{M}^{d+1}$, working at high energies and neglecting
particles masses. The scattering amplitude $\mathcal{A}$ is a
function of the Mandelstam invariants $s$, $t$, and can be
conveniently decomposed in \emph{S}--channel partial waves
\begin{equation*}
\mathcal{A=}s^{\frac{3-d}{2}}~\sum_{J\geq 0}\mathcal{S}_{J}\left(
z\right) ~e^{-2\pi i\,\sigma _{J}\left( s\right) }~,
\end{equation*}%
with
\begin{equation*}
z=\sin ^{2}\left( \frac{\theta }{2}\right) =-\frac{t}{s}
\end{equation*}%
related to the scattering angle $\theta $ and with $\sigma
_{J}\left(
s\right) $ the phase shift for the spin $J$ partial wave\footnote{%
We choose a non standard normalization and notation for the phase
shifts
for later convenience. To revert to standard conventions, one must replace $%
-2\pi \sigma _{J}\rightarrow 2\delta _{J}$.}. The angular functions $%
\mathcal{S}_{J}\left( z\right) $ are eigenfunctions of the
Laplacian on the sphere at infinity, with eigenvalue $-J\left(
J+d-2\right) $, and are polynomials in $z$ of order $J$, whose
explicit form depends on the dimension of spacetime. They can be
written as hypergeometric functions
\begin{equation*}
\mathcal{S}_{J}\left( z\right) \propto F\left. \left( -J,J+d-2,\frac{d-1}{2}%
\right\vert z\right) .
\end{equation*}%
and they are normalized so that $\sigma =0$ corresponds to free
propagation with no interactions. Unitarity then implies
$\mathrm{Im}\, \sigma \leq 0$. The amplitude itself can be
computed in perturbation theory
\begin{equation*}
\mathcal{A}=\mathcal{A}_{0}+\mathcal{A}_{1}+\cdots ~,
\end{equation*}%
where
$\mathcal{A}_{0}=s^{\frac{3-d}{2}}\sum_{J}\mathcal{S}_{J}\left(
z\right) ~$\ corresponds to graph $\left( a\right) $ in figure
\ref{fig4} describing free propagation in spacetime. All
\emph{S}--channel partial waves contribute to $\mathcal{A}_{0}$
with equal weight one, and vanishing phase shift.

In the eikonal regime we are interested in the limit $z\ll 1$ of
small scattering angle, where the amplitude is dominated by
partial waves with large intermediate spin $J$. One may then
replace the sum over $J$ with an integral over the impact
parameter $r$,
\begin{equation*}
\frac{r}{2}=\frac{J}{\sqrt{s}}~,
\end{equation*}%
denoting the phase shifts by $\sigma \left( s,r\right) $ from now
on. More precisely, if we consider the double limit
\begin{equation}
z\rightarrow0\, ,~\ \ \ \ \ \ J\rightarrow\infty\, ,~\ \ \ \ \ \ \
\ \ \ \ \ \ \left( z\sim J^{-2}\right),  \label{limitflat}
\end{equation}

\noindent we may approximate the sphere at infinity\ by a
transverse
euclidean plane $\mathbb{E}^{d-1}$ and the angular functions $\mathcal{S}%
_{J}\left( z\right) $ become, in this limit, the impact parameter
partial waves $\mathcal{I}_{J}$,

\begin{equation}
\mathcal{I}_{J}\propto \int_{\mathbb{E}^{d-1}}~dx~\delta \left(
x^{2}-r^{2}\right) ~e^{iq\cdot x}~\propto z^{\frac{3-d}{4}}\,\mathbf{J}_{%
\frac{d-3}{2}}\left( 2J\sqrt{z}\right) ,  \label{imprep2}
\end{equation}

\noindent with $q^{2}=-t$ and $\mathbf{J}_{\nu }$ the Bessel
function. One then obtains the impact parameter representation of
the amplitude
\begin{equation}
\mathcal{A}\,\simeq \,2s\int_{\mathbb{E}^{d-1}}~dx\;e^{iq\cdot
x}\;e^{-2\pi i\,\sigma \left( s,r\right) }\,,
\label{eikonalre--sum}
\end{equation}

\noindent where $r=\sqrt{x^{2}}$. In general, the phase shift
receives contributions at all orders in perturbation theory. On
the other hand, the leading behavior of $\sigma \left( s,r\right)
$ for large $r$, which controls small angle scattering, is
uniquely dominated by the leading tree--level interaction
$\mathcal{A}_{1}$ and it is therefore\ determined by a simple
Fourier transform

\begin{equation}
\mathcal{A}_{1}\,\simeq\,-4\pi
i~s\int_{\mathbb{E}^{d-1}}~dx\;e^{iq\cdot x}\;\sigma(s,r)\,.
\label{imprep1}
\end{equation}

\noindent The main result of the eikonal approximation is that the
knowledge of the tree--level amplitude $\mathcal{A}_{1}$ is enough
to compute the amplitude (\ref{eikonalre--sum}) in the
$z\rightarrow 0$ limit to all orders in the coupling. Moreover,
the dominant interaction in $\mathcal{A}_{1}$ comes from
\emph{T}--channel exchanges of spin $j$ massless particles, so
that the full amplitude (\ref{eikonalre--sum}) approximately
re--sums the crossed--ladder graphs in figure \ref{fig4}$(b)$. In
the limit of high energy $s$, the mediating particle with maximal
$j$ dominates the
interaction. In theories of gravity, this particle is the graviton, with $%
j=2 $.

\begin{figure}
[ptb]
\begin{center}
\includegraphics[height=1.3771in,width=4.7408in]
{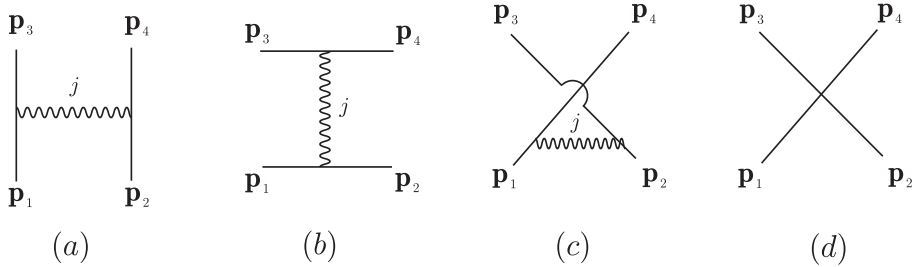} \caption{Some possible interactions at tree--level,
both in flat space and in $\mathrm{AdS}$. When decomposing the
amplitude in the $S$--channel, only graph $(a)$, with maximal spin
$j=2$, dominates the dynamics at large intermediate spin and
energy.} \label{fig2}
\end{center}
\end{figure}

To understand in more detail the generic behavior of amplitudes
for small scattering angle and large energies, consider some
sample interactions shown in figure \ref{fig2} contributing to
$\mathcal{A}_{1}$, where the exchanged particle has spin $j$.
The contribution of graph $2(a)$ to $\mathcal{A}_{1}$ has the form\footnote{%
The normalization $4^{3-j}~2\pi i\,G$ has been chosen for later
convenience. When $j=2$, the gravitational coupling constant $G$
is the canonically normalized Newton constant.}
\begin{equation*}
4^{3-j}~2\pi i\,G~\frac{s^{j}-c_{1}s^{j-1}t+\cdots}{-t}\propto
s^{j-1}\left( \frac{1}{z}+c_{1}+\cdots+c_{j}z^{j-1}\right) .
\end{equation*}
The polynomial part in $z$ contributes to partial waves with spin
$J<j$.
This can also be understood in the impact parameter representation (\ref%
{imprep1}), since polynomial terms in $z=q^{2}s$ give, after
Fourier
transform, delta function contributions to the phase shift localized at $%
r=0$. The universal term $s^{j-1}/z$ contributes, on the other
hand, to partial waves of all spins and gives a phase shift at
large $r$ given by

\begin{equation}
\sigma\left( s,r\right) \simeq-8G\left( \frac{s}{4}\right) ^{j-1}
\Pi\left( r\right) ~,  \label{phaseshift}
\end{equation}

\noindent where $\Pi\left( r\right)$ is the massless Euclidean
propagator in transverse space $\mathbb{E}^{d-1}$. At high
energies, the maximal $j$
dominates. Graph $2(b)$ is obtained by interchanging the role of $s$ and $t$%
, and therefore is proportional to $s^{j-1}\left( z^{j}+c_{1}
z^{j-1}+\cdots+c_{j}\right)$. The corresponding intermediate
partial waves have spin $J\leq j$. Finally, graph $2(c)$ is
obtained from $2(a)$ by
replacing $t$ by $u=-\left( s+t\right) $ and therefore by sending $%
z\rightarrow 1-z$. Using the fact that $\mathcal{S}_{J}(1-z)=(-)^J\mathcal{S}%
_{J}(z)$, we can again expand the \emph{U}--channel exchange graph
$2(c)$ in
the form $s^{\frac{3-d}{2}}\sum_{J}\left( -\right) ^{J}\mathcal{S}%
_{J}~\sigma_{J}\left( s\right)$ where $\sigma_J(s)$ are the phase
shifts of
the \emph{T}--channel exchange, whose large spin behavior is given by (\ref%
{phaseshift}). At large spins, the contribution to the various
partial waves is alternating in sign and averages to a
sub--leading contribution. Finally, the contact graph $2(d)$ only
contains a spin zero
contribution. We conclude that, at large spins and energies, the \emph{T}%
--channel exchange of a graviton in graph $2(a)$ dominates all
other interactions.

We shall study the CFT analogue of $2\rightarrow 2$ scattering of
scalars in flat space. More precisely, we will study the CFT
correlator
\begin{equation*}
\left\langle \mathcal{O}_{1}\left( \mathbf{p}_{1}\right) \mathcal{O}%
_{2}\left( \mathbf{p}_{2}\right) \mathcal{O}_{1}\left(
\mathbf{p}_{3}\right)
\mathcal{O}_{2}\left( \mathbf{p}_{4}\right) \right\rangle _{\text{\textrm{CFT%
}}_{d}}\equiv \frac{1}{\mathbf{p}_{13}^{\Delta
_{1}}\mathbf{p}_{24}^{\Delta _{2}}}\,\, \mathcal{A}\left(
z,\bar{z}\right) ~,
\end{equation*}%
where the scalar primary operators $\mathcal{O}_{1}$,
$\mathcal{O}_{2}$ have
conformal dimensions $\Delta _{1}$, $\Delta _{2}$, respectively. The $%
\mathbf{p}_{i}$ are now points on the boundary of $\mathrm{AdS}$,
and the
amplitude $\mathcal{A}$ is now a function of two cross--ratios $z$, $\bar{z}$%
, whose precise definition is given in section \ref{notation}.
Neglecting
string effects, the function $\mathcal{A}$ can be computed in the dual $%
\mathrm{AdS}$ formulation as a field theoretic perturbation series
in the gravitational coupling \cite{WittenGubser, w98, fmmr98,
Rastelli1, dmmr99}
\begin{equation*}
\mathcal{A}=\mathcal{A}_{0}+\mathcal{A}_{1}+\cdots ~,
\end{equation*}%
where $\mathcal{A}_{0}=1$ corresponds to free propagation
described again in figure \ref{fig4}$(a)$. The amplitude
$\mathcal{A}$ can be decomposed, as in flat space, in
\emph{S}--channel partial waves. These are know, in the CFT
litterature, as conformal partial waves and correspond to the
exchange of
the conformal primaries that appear in the operator product expansion (OPE) of $\mathcal{O}_{1}$ with $%
\mathcal{O}_{2}$ as $\mathbf{p}_{1}\rightarrow \mathbf{p}_{2}$,
together with their descendants. In particular, the free amplitude
$\mathcal{A}_{0}$ corresponds to the exchange of an infinite set
of $\mathcal{O}_{1}\partial _{\mu _{1}}\cdots \partial _{\mu
_{J}}\partial ^{2n}\mathcal{O}_{2}$
composites\footnote{%
Throughout this paper we will use this schematic notation to
represent the primary composite operators of spin $J$ and
conformal dimension $E=\Delta _{1}+\Delta _{2}+J+2n$, avoiding the
rather cumbersome exact expression. We
shall also use the simpler notation $\mathcal{O}_{1}\partial ^{N}\mathcal{O}%
_{2}$ whenever possible.} of spin $J$ and dimension $E=\Delta
_{1}+\Delta _{2}+J+2n$, with $J,n$ non--negative integers.

As the coupling is turned on, we expect that the above lattice of
intermediate primaries acquires anomalous dimensions, and also
that new intermediate states appear in the partial wave
decomposition. We shall show that the large spin and dimension
\emph{S}--channel decomposition of the tree level amplitude
$\mathcal{A}_{1}$ is dominated, as in flat space, by the
\emph{T}--channel exchange of massless particles of maximal spin
$j$. In
fact, in this limit, this decomposition is determined by the small $z$, $%
\bar{z}$ behavior of the discontinuity across a kinematical branch
cut of the \emph{T}--channel exchange Witten diagram
\ref{fig2}$(a)$, which we derived in \cite{ourSW}. This result is
central in the derivation of our main findings:

\begin{itemize}
\item[$(i)$] The AdS graph \ref{fig2}$(a)$ contributes to all
partial waves corresponding to the
$\mathcal{O}_{1}\partial^{N}\mathcal{O}_{2}$ composites of spin
$J$ and dimension $E$ already present in $\mathcal{A}_{0}$. For
large $E$, $J$ these composites acquire anomalous dimensions given
by
\begin{equation*}
2\sigma\left( s,r\right) ~,
\end{equation*}
where $\sigma\left( s,r\right) $ is again given by
(\ref{phaseshift}), but where now
\begin{align*}
s& =\left( E^{2}-J^{2}\right) /\ell ^{2}~, \\
\tanh \left( \frac{r}{2\ell }\right) & =\frac{J}{E}~,
\end{align*}%
and where $\Pi \left( r\right) $ is the \textit{massive} Euclidean
propagator on transverse space, which is now the hyperbolic space $\mathrm{H}%
_{d-1}$ with radius $\ell $. The mass--squared is given by $\left(
d-1\right) /\ell ^{2}$ and $r$ is the geodesic distance on $\mathrm{H}_{d-1}$%
. Once again, the dominant contribution to $\sigma $ comes from
the exchange
of the graviton. Note that the flat space limit can be obtained letting $%
\ell \rightarrow \infty $ keeping the physical
$\mathrm{AdS}$\emph{\ }energy $E/\ell $ fixed. (Sections
\ref{TreeS} and \ref{GravDom})

\item[$(ii)$] The AdS graph \ref{fig2}$(b)$ contributes to partial
waves
with spin $J\leq j$ and with dimensions $E=d$, $E=\Delta _{1}+\Delta _{2}+n$%
. These partial waves correspond to the massless exchanged
particle and to corrections to the composites
$\mathcal{O}_{1}\partial ^{N}\mathcal{O}_{2}$ with bounded spin.
Moreover, the contribution with maximal spin $J=j$ can be
determined explicitly without computing the full graph. (Sections
\ref{TreeT} and \ref{GravDom})

\item[$(iii)$] The AdS graph \ref{fig2}$(c)$ is similar to graph \ref{fig2}$%
(a)$, but with alternating signs $\left( -\right) ^{J}$. Graph \ref{fig2}$%
(d) $ contributes to partial waves with spin $J=0$ only. Comments
similar to the flat space case apply. (Section \ref{GravDom})

\item[$(iv)$] In order to arrive at the above results, we have
extended the impact parameter representations (\ref{imprep2}) and
(\ref{imprep1}) to the decomposition of $\mathrm{CFT}$ amplitudes
in terms of conformal
partial waves, using the approach of \cite{Osborn, Osborn22}. (Section \ref%
{impactsection})
\end{itemize}

Finally, in section \ref{Conj}, we conjecture a formula for
$\mathcal{A}$ which resums the perturbative expansion in the AdS
gravitational coupling $G$ in the eikonal limit. This formula also
predicts the anomalous dimensions of $\mathcal{O}_{1}\partial
^{N}\mathcal{O}_{2}$ composites with large spin and conformal
dimension to all orders in $G$.


\section{Preliminaries and Notation\label{notation}}


This section follows closely section 2 of \cite{ourSW} and is
included for completeness. Recall that \textrm{AdS}$_{d+1}$ space,
of dimension $d+1$, can be defined as a pseudo--sphere in the
embedding space $\mathbb{M}^{2} \times\mathbb{M}^{d}$. We denote
with $\mathbf{x}=\left( x^{+},x^{-} ,x\right)  $ a point in
$\mathbb{M}^{2}\times\mathbb{M}^{d}$, where $x^{\pm}$ are
light--cone coordinates on $\mathbb{M}^{2}$ and $x$ denotes a
point in $\mathbb{M}^{d}$. Then, the AdS space of radius $\ell$ is
described by\footnote{We denote with $\mathbf{x}\cdot\mathbf{y}$
and $x\cdot y$ the scalar products in
$\mathbb{M}^{2}\times\mathbb{M}^{d}$ and $\mathbb{M}^{d}$,
respectively. Moreover we abbreviate
$\mathbf{x}^{2}\mathbf{=x}\cdot \mathbf{x}$ and $x^{2}=x\cdot x$
when clear from context. In $\mathbb{M}^{d}$ we shall use
coordinates $x^\mu$ with $\mu=0,\ldots,d-1$ and with $x^0$ the
timelike coordinate.}
\begin{equation}
\mathbf{x}^{2}=-x^{+}x^{-}+x^{2}=-\ell^{2}~.\label{pseudoS}
\end{equation}
\noindent
Similarly, a point on the holographic boundary of \textrm{AdS}$_{d+1}$ can be
described by a ray on the light--cone in $\mathbb{M}^{2}\times\mathbb{M}^{d}$,
that is by a point $\mathbf{p}$ with
\[
\mathbf{p}^{2}=-p^{+}p^{-}+p^{2}=0~,
\]
defined up to re--scaling
\[
\mathbf{p}\sim\lambda
\mathbf{p} \qquad \left(
\lambda>0\right).
\]
From now on we choose units such that $\ell=1$.

The AdS/CFT correspondence predicts the existence of a dual CFT$_{d}$ living
on the boundary of AdS$_{d+1}$. In particular, a CFT correlator of scalar primary operators located at points $\mathbf{p}_{1},\ldots,\mathbf{p}_{n}$ can be conveniently described by an amplitude
\[
A\left(  \mathbf{p}_{1},\ldots,\mathbf{p}_{n}\right)
\]
invariant under $SO\left(  2,d\right)  $ and therefore only a function of the
invariants
\[
\mathbf{p}_{ij}=-2\mathbf{p}_{i}\cdot\mathbf{p}_{j}~.
\]
Since the boundary points $\mathbf{p}_{i}$ are defined only up to re--scaling,
the amplitude $A$ will be homogeneous in each entry
\[
A\left(  \ldots,\lambda\mathbf{p}_{i},\ldots\right)  =\lambda^{-\Delta_{i}
}A\left(  \ldots,\mathbf{p}_{i},\ldots\right)  ~,
\]
where $\Delta_{i}$ is the conformal dimension of the $i$--th scalar primary operator.

Throughout this paper we will focus our attention on four--point amplitudes of
scalar primary operators. More precisely, we shall consider correlators of the form
\[
A\left(  \mathbf{p}_{1},\mathbf{p}_{2},\mathbf{p}_{3},\mathbf{p}_{4}\right)
=\left\langle \mathcal{O}_{1}\left(  \mathbf{p}_{1}\right)  \mathcal{O}
_{2}\left(  \mathbf{p}_{2}\right)  \mathcal{O}_{1}\left(  \mathbf{p}
_{3}\right)  \mathcal{O}_{2}\left(  \mathbf{p}_{4}\right)  \right\rangle
_{\text{\textrm{CFT}}_{d}}
\]
where the scalar operators $\mathcal{O}_{1},\mathcal{O}_{2}$ have dimensions
\[
\Delta_{1}=\Delta+\nu~,~\ \ \ \ \ \ \ \ \ \ \ \ \ \ \ \ \ \ \ \Delta
_{2}=\Delta-\nu~,
\]
respectively. The four--point amplitude $A$ is just a function of
two cross ratios $z,\bar{z}$ which we define, following
\cite{Osborn, Osborn22}, in terms of the kinematical invariants
$\mathbf{p}_{ij}$ as\footnote{Throughout the paper, we shall
consider barred and unbarred variables as independent, with
complex conjugation denoted by $\star$. In general
$\bar{z}=z^{\star}$ when considering the analytic continuation of
the CFT$_{d}$ to the Euclidean signature.}
\begin{align*}
z\bar{z}  & =\frac{\mathbf{p}_{13}\mathbf{p}_{24}}{\mathbf{p}_{12}
\mathbf{p}_{34}}~,\\
\left(  1-z\right)  \left(  1-\bar{z}\right)   & =\frac{\mathbf{p}
_{14}\mathbf{p}_{23}}{\mathbf{p}_{12}\mathbf{p}_{34}}~.
\end{align*}
\noindent
Then, the four--point amplitude can be written as
\[
A\left(  \mathbf{p}_{i}\right)
=\frac{1}{\mathbf{p}_{13}^{\Delta_{1}
}\mathbf{p}_{24}^{\Delta_{2}}}\,\,\mathcal{A}\left(
z,\bar{z}\right) ~,
\]
where $\mathcal{A}$ is a generic function of $z,\bar{z}$. By
conformal invariance, we can fix the position of up to three of
the external points $\mathbf{p}_{i}$. In what follows, we shall
often choose the external kinematics by placing the four points
$\mathbf{p}_{i}$ at
\begin{align}
\mathbf{p}_{1}  & =\left(  0,1,0\right)
~,\ \ \ \ \ \ \ \ \ \ \ \ \ \ \ \ \ \ \ \ \ \ \ \mathbf{p}_{2}=-\left(
1,p^{2},p\right)  ~,\label{extpoints}\\
\mathbf{p}_{3}  & =-\left(  q^{2},1,q\right)
~,\ \ \ \ \ \ \ \ \ \ \ \ \ \ \ \ \ \ \ \mathbf{p}_{4}=\left(  1,0,0\right)
~,\nonumber
\end{align}
\noindent and we shall view the amplitude as a function of
$p,q\in\mathbb{M}^{d}$. The cross ratios $z,\bar{z}$ are in
particular determined by
\[
z\bar{z}=q^{2}p^{2},~\ \ \ \ \ \ \ \ \ \ \ \ \ \ \ \ z+\bar{z}=2p\cdot q~.
\]
When $d=2$ it is convenient to parameterize $\mathbb{M}^{d}$ by light--cone
coordinates $x=x^0+x^1$, $\bar{x}=x^0-x^1$ with metric $-dxd\bar{x}$. Then, if we choose $p=\bar{p}=-1$ we have $q=z$, $\bar{q}=\bar{z}$.

In the sequel, we will denote with
$\mathrm{H}_{d-1}\subset\mathbb{M}^{d}$ the transverse hyperbolic
space, given by the upper mass--shell
\[
x^{2}=-1~,\ \ \ \ \ \ \ \ \ \ \ \ \ \ \left(  x^{0}>0\right)
\]
where $x\in\mathbb{M}^{d}$. We will also denote with
$\mathrm{M}\subset\mathbb{M}^{d}$ the future Milne wedge given by
$x^{2}\leq0$, \thinspace$x^{0}\geq0$. Similarly, we denote with
$-\mathrm{M}$ the past Milne wedge and with $-\mathrm{H}_{d-1}$
the corresponding transverse hyperbolic space. Finally we denote
with
\begin{equation*}
\widetilde{dx}   =2dx~\delta\left(  x^{2}+1\right)  ,
\end{equation*}
\noindent the volume element on $\mathrm{H}_{d-1}$, such that
\[
\int_{\mathrm{M}}dx = \int_{\mathrm{M}}d(ty) \int_{0}^{\infty}t^{d-1}dt\int_{\mathrm{H}_{d-1} }\widetilde{dy}~,
\]
where $x=ty$ and $y\in \mathrm{H}_{d-1}$.

Throughout the paper, we will often need the massive Euclidean
scalar propagator $\Pi\left( x,y\right)$ on $\mathrm{H}_{d-1}$, of
mass--squared $d-1\,$, defined by
\begin{equation*}
\left[  \square_{\mathrm{H}_{d-1}}-\left(  d-1\right)  \right]
\,\Pi\left( x,y\right)    =-~\delta\left(  x,y\right)  ~,
\end{equation*}
\noindent and explicitly given in terms of the hypergeometric
function
\begin{equation}
\Pi\left(  x,y\right)  =\left(  4\pi\right)  ^{\frac{1-d}{2}}
\frac{  \Gamma\left(  \frac{d+1}{2}\right) }{d(d-1) }\left(
-z\right)  ^{1-d} F \left. \left(  d-1,\frac{d+1} {2},d+1 \right|
\frac{1}{z}\right)  \ , \label{explicitprop}
\end{equation}

\noindent
where $2z=1+x\cdot y$.


\section{Conformal Partial Waves}

\begin{figure}
[ptb]
\begin{center}
\includegraphics[height=1.8449in,width=2.9107in]
{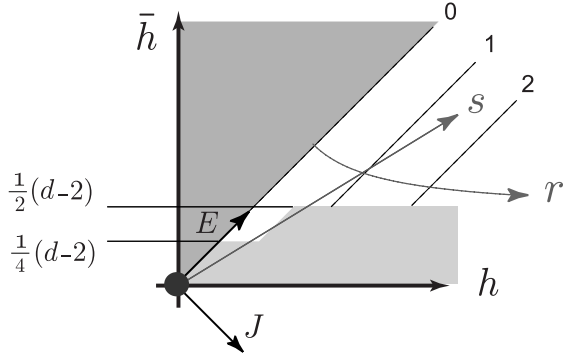} \caption{Intermediate primaries of dimension
$E=h+\bar{h}$ and integer spin $J=h-\bar{h}\ge 0$ can exist in the
white region along the dashed lines. The light grey region is
excluded due to the unitarity constraint, with the vacuum at
$h=\bar{h}=0$ being the unique exception. Also shown are the
impact parameter labels $s=4h\bar{h}$ and $r=\ln\left(
h/\bar{h}\right)$.} \label{fig3}
\end{center}
\end{figure}

The amplitude $\mathcal{A}\left(  z,\bar{z}\right)  $ can be
expanded using the OPE around $z,\bar{z}=0,1,\infty$,
corresponding to the point $\mathbf{p}_{3}$ getting close to
$\mathbf{p}_{1}$, $\mathbf{p}_{2}$ and $\mathbf{p}_{4}$,
respectively. In particular, we will be interested in the
contribution to the amplitude $\mathcal{A}$ coming from the
exchange of a conformal primary operator of dimension $E$ and
integer spin $J\geq0$ in the two channels
\begin{align*}
z,\bar{z}  & \rightarrow
0\ ,\ \ \ \ \ \ \ \ \ \ \ \ \ \ \ \ \ \ \ \ \ \ \text{\emph{T}--channel},\\
z,\bar{z}  & \rightarrow\infty
\ ,\ \ \ \ \ \ \ \ \ \ \ \ \ \ \ \ \ \ \ \ \ \text{\emph{S}--channel},
\end{align*}
\noindent together with all of its conformal descendants. It will
be convenient in the following to use different labels for energy
and spin $E,J$. We shall use most frequently conformal dimensions
$h,\bar{h}$ defined by
\begin{align*}
E  & =h+\bar{h}~,\\
J  & =h-\bar{h}~.
\end{align*}
\noindent We will also use the so called impact parameter labels
$s,r$ defined by
\begin{align}
s  & =4h\bar{h}=E^{2}-J^{2}~,\nonumber\\
e^{-r}  & =\frac{\bar{h}}{h}=\frac{E-J}{E+J}~.\label{ImpactParam}
\end{align}
\noindent In terms of these variables, the present equations
closely resemble eikonal results in flat space, with $s$ playing
the role of the total center--of--mass energy squared, and with
$r$ being the physical transverse impact parameter. We shall
justify this interpretation more clearly in section
\ref{SectFree}. We also recall the unitarity bounds $E\geq d-2+J$
for $J\geq1$ and $E\geq\left( d-2\right)  /2$ for $J=0$, with the
single exception of the vacuum with $E=J=0$. This translates into
\begin{align*}
\bar{h}  & \geq\frac{d-2}{2}~,\ \ \ \ \ \ \ \ \ \ \ \ \ \ \ \ \ \ \ \ \left(
J\geq1\right), \\
\bar{h}  & \geq\frac{d-2}{4}~,\ \ \ \ \ \ \ \ \ \ \ \ \ \ \ \ \ \ \ \ \left(
J=0\right),
\end{align*}
\noindent
again with the exception of the vacuum at $h=\bar{h}=0$. Figure \ref{fig3}
summarizes the basic notation regarding the intermediate conformal primaries.

The amplitude $\mathcal{A}\left(  z,\bar{z}\right)  $ can be
expanded in the basis of conformal partial waves in either the
\emph{T} or \emph{S}--channels, whose elements we shall denote by
\[
\mathcal{T}_{h,\bar{h}}~\left(  z,\bar{z}\right)
~,\ \ \ \ \ \ \ \ \ \ \ \ \ \ \ \ \ \ \ \ \mathcal{S}_{h,\bar{h}}~\left(
z,\bar{z}\right)  ~.
\]
Following \cite{Osborn, Osborn22}, the functions
$\mathcal{T}_{h,\bar{h}}$ and $\mathcal{S}_{h,\bar{h}}$ must be
symmetric in $z,\bar{z}$ and must satisfy the differential
equations
\begin{align*}
D_{T}~\mathcal{T}_{h,\bar{h}}=c_{h,\bar{h}}~\mathcal{T}_{h,\bar{h}}~,  & \\
\left(  z\bar{z}\right)  ^{\Delta}D_{S}~\left[  \left(  z\bar{z}\right)
^{-\Delta}\mathcal{S}_{h,\bar{h}}\right]  =c_{h,\bar{h}}~\mathcal{S}
_{h,\bar{h}}~,  &
\end{align*}
\noindent where the constant $c_{h,\bar{h}}$ is the Casimir of the
conformal group given by
\[
c_{h,\bar{h}}=h\left(  h-1\right)  +\bar{h}\left(  \bar{h}-d+1\right),
\]
and where the differential operators $D_{T},~D_{S}$ have the
explicit form
\begin{align}
D_{T}  & =z^{2}\left(  1-z\right)  \partial^{2}-z^{2}\partial+\bar{z}
^{2}\left(  1-\bar{z}\right)  \bar{\partial}^{2}-\bar{z}^{2}\bar{\partial
} +\label{Tchannel}\\
& +\left(  d-2\right)  \frac{z\bar{z}}{z-\bar{z}}\left[  \left(  1-z\right)
\partial-\left(  1-\bar{z}\right)  \bar{\partial}\right] \nonumber
\end{align}
\noindent and
\begin{align}
D_{S}  & =z\left(  z-1\right)  \partial^{2}+\left(  2z-1\right)
\partial+\frac{\nu^{2}}{z}+\nonumber\\
& +\bar{z}\left(  \bar{z}-1\right)  \bar{\partial}^{2}+\left(  2\bar
{z}-1\right)  \bar{\partial}+\frac{\nu^{2}}{\bar{z}}+\label{Schannel}\\
& +\frac{d-2}{z-\bar{z}}\left[  z\left(  z-1\right)  \partial-\bar{z}\left(
\bar{z}-1\right)  \bar{\partial}\right]  ~.\nonumber
\end{align}
\noindent Moreover, the partial waves $\mathcal{T}_{h,\bar{h}}$
and $\mathcal{S} _{h,\bar{h}}$ satisfy the boundary conditions
\begin{align*}
\lim_{z,\bar{z}\rightarrow0}\mathcal{T}_{h,\bar{h}}  & \sim z^{h}~\bar
{z}^{\bar{h}}~,\\
\lim_{z,\bar{z}\rightarrow\infty}\mathcal{S}_{h,\bar{h}}  & \sim z^{\Delta
-h}~\bar{z}^{\Delta-\bar{h}}~,
\end{align*}
\noindent where we choose to take the limit
$\bar{z}\rightarrow0,\infty$ first. The symmetric term with $h$
and $\bar{h}$ interchanged is then sub--leading since
$h\geq\bar{h}$.


\subsection{The $d=2$ Case}


Explicit expressions for the partial waves
$\mathcal{T}_{h,\bar{h}}$ and $\mathcal{S}_{h,\bar{h}}$ exist for
$d$ even \cite{Osborn, Osborn22}, and are particularly simple in
$d=2$ where the problem factorizes in left/right equations for $z$
and $\bar{z}$. In this case we have the explicit expressions
\begin{align}
\mathcal{T}_{h,\bar{h}}\left(  z,\bar{z}\right)   & =\mathcal{T}_{h}\left(
z\right)  \mathcal{T}_{\bar{h}}\left(  \bar{z}\right)  +\mathcal{T}_{\bar{h}
}\left(  z\right)  \mathcal{T}_{h}\left(  \bar{z}\right)  ~,\label{eq1000}\\
\mathcal{S}_{h,\bar{h}}\left(  z,\bar{z}\right)   & =\mathcal{S}_{h}\left(
z\right)  \mathcal{S}_{\bar{h}}\left(  \bar{z}\right)  +\mathcal{S}_{\bar{h}
}\left(  z\right)  \mathcal{S}_{h}\left(  \bar{z}\right)  ~,\nonumber
\end{align}
\noindent for $h>\bar{h}$ and
\begin{align*}
\mathcal{T}_{h,h}\left(  z,\bar{z}\right)   & =\mathcal{T}_{h}\left(
z\right)  \mathcal{T}_{h}\left(  \bar{z}\right)  ~, \\
\mathcal{S}_{h,h}\left(  z,\bar{z}\right)   & =\mathcal{S}_{h}\left(
z\right)  \mathcal{S}_{h}\left(  \bar{z}\right)  ~,
\end{align*}
\noindent for $h=\bar{h}$, where
\begin{align}
\mathcal{T}_{h}\left(  z\right)   & =\left(  -z\right)  ^{h}~F \Big(
h,h,2h \Big| z\Big)  ~,\nonumber \\
\mathcal{S}_{h}\left(  z\right)   & =a_{h}~\left(  -z\right)
^{\Delta -h}~F\Big(  h+\nu,h-\nu,2h \Big| z^{-1}\Big)
~.\label{2dCPW}
\end{align}

\noindent The specific normalization of the \emph{S}--channel
partial waves
\[
a_{h}=\frac{\Gamma\left(  h+\nu\right)  \Gamma\left(  h-\nu\right)
\Gamma\left(  h+\Delta-1\right)  }{\Gamma\left(  \Delta+\nu\right)
\Gamma\left(  \Delta-\nu\right)  \Gamma\left(  2h-1\right)  \Gamma\left(
h-\Delta+1\right)  }
\]
is chosen for later convenience, and it is such that
\begin{equation}
\sum_{h\in\Delta+\mathbb{N}_{0}}\mathcal{S}_{h}\left(  z\right)
=1~,\label{ChiralSum}
\end{equation}
\noindent
where $\mathbb{N}_{0}$ is the set of non--negative integers.

It is clear from (\ref{eq1000}) that the $h,\bar{h}$ partial waves
correspond to the exchange of a pair of primary operators of
holomorphic/antiholomorphic dimension $\left(  h,\bar{h}\right)  $
and $\left(  \bar{h},h\right)  $, together with their descendants.


\subsection{Impact Parameter Representation\label{impactsection}}


We now move back to general dimension $d$, and we consider the
behavior of the \emph{S}--channel partial waves
$\mathcal{S}_{h,\bar{h}}\left(  z,\bar{z}\right)$ for $z,\,
\bar{z} \to 0$, \textit{i.e.}, in the dual \emph{T}--channel. More
precisely, in strict analogy with the case of flat space, we
analyze the double limit
\begin{align*}
z,\bar{z}  & \rightarrow0~,\\
h,\bar{h}  & \rightarrow\infty~,
\end{align*}
\noindent
as in (\ref{limitflat}), with
\[
z \sim \bar{z}\sim h^{-2} \sim \bar{h}^{-2}~.
\]
In this limit, the differential operator $D_{S}$ in (\ref{Schannel}) and the
constant $c_{h,\bar{h}}$ reduce to
\[
-\tilde{D}_{S}=z\partial^{2}+\bar{z}\bar{\partial}^{2}+\partial+\bar{\partial
}-\frac{\nu^{2}}{z}-\frac{\nu^{2}}{\bar{z}}+\frac{d-2}{z-\bar{z}}\left(
z\partial-\bar{z}\bar{\partial}\right)  ~
\]
and
\[
\tilde{c}_{h,\bar{h}}=h^{2}+\bar{h}^{2}~.
\]
We shall denote with $\mathcal{I}_{h,\bar{h}}\left(
z,\bar{z}\right)  $ the approximate \textit{impact parameter}
\emph{S}--channel partial wave, which satisfies

\begin{equation}
\left(  -\tilde{D}_{S}+\tilde{c}_{h,\bar{h}}\right)  \left[  \left(  z\bar
{z}\right)  ^{-\Delta}\mathcal{I}_{h,\bar{h}}\right]  =0~.
\label{PDEimpact}
\end{equation}

\noindent Fixing the external points $\mathbf{p}_{i}$ as in
(\ref{extpoints}), we view the \emph{S}--channel impact parameter
amplitude $\mathcal{I}_{h,\bar{h}}$ as a function of $q$ and $p$.
In analogy with the flat space case (\ref{imprep2}), the function\
$\mathcal{I}_{h,\bar{h}}$ admits the following integral
representation over the future Milne wedge $M$

\begin{align}
\mathcal{I}_{h,\bar{h}}  & =\mathcal{N}_{\Delta_{1}}\mathcal{N}_{\Delta_{2}
}~\left(  -q^{2}\right)  ^{\Delta_{1}}\left(  -p^{2}\right)  ^{\Delta_{2}}
\int_{\mathrm{M}}\frac{dx}{\left\vert x\right\vert ^{d-2\Delta_{1}}}\frac
{dy}{\left\vert y\right\vert ^{d-2\Delta_{2}}}~~e^{-2q\cdot x-2p\cdot
y}~\times\nonumber\\
& \times~4h\bar{h}\left(  h^{2}-\bar{h}^{2}\right)  ~\delta\left(  2x\cdot
y~+h^{2}+\bar{h}^{2}\right)  ~\delta\left(  x^{2}y^{2}-h^{2}\bar{h}
^{2}\right)  ~,\label{eq1001}
\end{align}
\noindent
where

$$
\mathcal{N}_{\Delta}^{-1} = \int_{\mathrm{M}}\frac{dy}{\left\vert
y\right\vert ^{d-2\Delta}}~~e^{2k\cdot y}=\Gamma\left(
2\Delta\right) \int_{\mathrm{H}_{d-1}}\frac{\widetilde{dy}}{\left(
-2k\cdot y\right) ^{2\Delta}} \frac{\pi^{\frac{d}{2}-1}}{2}\Gamma\left(  \Delta\right)
\Gamma\left( \Delta-\frac{d}{2}+1\right)  ~,
$$
with $k\in\mathrm{H}_{d-1}$ arbitrary\footnote{In the appendix, we
show that (\ref{eq1001}) is a solution of the differential
equation (\ref{PDEimpact}). On the other hand, we do not have a
complete proof of (\ref{eq1001}), since we cannot distinguish
different cases with the same Casimir $\tilde{c}_{h,\bar{h}}$. On
the other hand, the form (\ref{eq1001}) is strongly suggested by
the case $d=2$, where we can check explicitly that (\ref{eq1001})
is the impact parameter approximation to $\mathcal{S}_{h,\bar{h}}$
(see section \ref{d=2Impact})}. Any function $\Sigma(z,\bar{z})$
can be decomposed in the impact parameter partial waves
$\mathcal{I} _{h,\bar{h}}$ and we have chosen the normalization of
$\mathcal{I}_{h,\bar{h} }$ such that
\begin{align}
 \Sigma\left(  z,\bar{z}\right) & =\int_{0}^{\infty}dh\int_{0}^{h}d\bar
{h}~~\sigma\left(  h^{2}\bar{h}^{2},h^{2}+\bar{h}^{2}\right)  ~~\mathcal{I}
_{h,\bar{h}}\left(  z,\bar{z}\right) \label{intrep}\\
&=\mathcal{N}_{\Delta_{1}}\mathcal{N}_{\Delta_{2}}~\left(  -q^{2}\right)
^{\Delta_{1}}\left(  -p^{2}\right)  ^{\Delta_{2}}\times\nonumber\\
&\ \ \  \times\int_{\mathrm{M}}\frac{dx}{\left\vert x\right\vert ^{d-2\Delta_{1}}
}\frac{dy}{\left\vert y\right\vert ^{d-2\Delta_{2}}}~~e^{-2q\cdot x-2p\cdot
y}~\sigma\left(  x^{2}y^{2},-2x\cdot y\right)  ~.\nonumber
\end{align}
\noindent
In particular, setting $\sigma=1$ we get
\[
\int_{0}^{\infty}dh\int_{0}^{h}d\bar{h}~~~\mathcal{I}_{h,\bar{h}}=1~.
\]
Note that, in (\ref{intrep}), the leading behavior of the function
$\Sigma$ for $z,\bar{z}\rightarrow0$ is controlled by the behavior
of $\sigma$ for $h,\bar{h}\rightarrow\infty$. In fact, when
$\sigma\sim\left( h\bar{h}\right)^{-a}$ for large $h,\bar{h}$ with
$a<2\Delta_1$ and $a<2\Delta_2$, then $\Sigma\sim\left(
z\bar{z}\right)^{a/2}$ for small $z,\bar{z}$.


\subsection{Impact Parameter Representation in $d=2$} \label{d=2Impact}


In $d=2$ the constant $\mathcal{N}_{\Delta}$ is given explicitly
by $2/\Gamma\left(  \Delta\right)  ^{2}$. Choosing $p=\bar{p}=-1$
and $q=z,$ $\bar{q}=\bar{z}$, the general expression
(\ref{eq1001}) reduces to

\begin{align*}
\mathcal{I}_{h,\bar{h}}  & =\frac{\left(  z\bar{z}\right)  ^{\Delta_{1}}
}{\Gamma\left(  \Delta_{1}\right)  ^{2}\Gamma\left(  \Delta_{2}\right)  ^{2}
}~\int_{0}^{\infty}\frac{dxd\bar{x}}{\left(  x\bar{x}\right)  ^{1-\Delta_{1}}
}\frac{dyd\bar{y}}{\left(  y\bar{y}\right)  ^{1-\Delta_{2}}}~~e^{z\bar
{x}+z\bar{x}-y-\bar{y}}\times\\
& \times4h\bar{h}\left(  h^{2}-\bar{h}^{2}\right)  ~\delta\left(  x\bar
{y}+y\bar{x}~-h^{2}-\bar{h}^{2}\right)  ~\delta\left(  x\bar{x}y\bar{y}
-h^{2}\bar{h}^{2}\right)  ~.
\end{align*}

\noindent
The $y,\bar{y}$ integrals localize at two points, namely at
\[
y=\bar{x}^{-1}h^{2},~\ \ \ \ \ \ \ \ \ \ \ \ \ \ \bar{y}=x^{-1}\bar{h}^{2}~,
\]
and at the point obtained by exchanging $h$ with $\bar{h}$.
Summing the two contributions we obtain
\[
\mathcal{I}_{h,\bar{h}}\left(  z,\bar{z}\right)  =\mathcal{I}_{h}\left(
z\right)  \mathcal{I}_{\bar{h}}\left(  \bar{z}\right)  +\mathcal{I}_{\bar{h}
}\left(  z\right)  \mathcal{I}_{h}\left(  \bar{z}\right)  ~,
\]
where
\begin{align*}
\mathcal{I}_{h}\left(  z\right)   & =2\, \frac{h^{2\Delta_{2}-1}~\left(
-z\right)  ^{\Delta_{1}}}{\Gamma\left(  \Delta_{1}\right)  \Gamma\left(
\Delta_{2}\right)  }\int_{0}^{\infty}\frac{d\bar{x}}{\bar{x}^{1-2\nu}
}~~e^{z\bar{x}-h^{2}\bar{x}^{-1}}\\
& =4\, \frac{h^{2\Delta-1}\left(  -z\right)  ^{\Delta}}{\Gamma\left(  \Delta
_{1}\right)  \Gamma\left(  \Delta_{2}\right)  }~~K_{2\nu}\left(  2h\sqrt
{-z}\right)  ~.
\end{align*}

\noindent One can check that the function $\mathcal{I}_{h}\left(
z\right)$ is indeed the impact parameter approximation of
$\mathcal{S}_{h}\left(z\right)$ in (\ref{2dCPW}). Moreover it
satisfies
\[
\int_{0}^{\infty}dh~\mathcal{I}_{h}\left(  z\right)  =1~,
\]
which corresponds to (\ref{ChiralSum}).


\section{Propagation in AdS{}\label{SecPropagation}}

\begin{figure}
[ptb]
\begin{center}
\includegraphics[height=1.457in,width=3.1306in]
{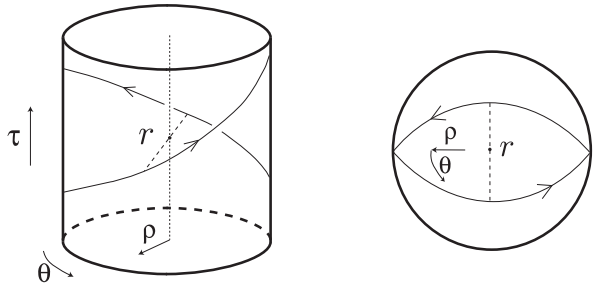}
\caption{Collision of two particles in AdS. The trajectories are shown both in
the full space $\tau,\rho,\theta$ as well as projected onto a spatial slice
$\rho,\theta$. The full configuration has total energy $E$ and spin $J$, with
the particles reaching a minimum geodesic distance $r$ given by $\tanh\left( r/2\right)=J/E$.}
\label{fig5}
\end{center}
\end{figure}

The impact parameter $r$ in (\ref{ImpactParam}) has a natural
interpretation in the dual AdS geometry. Writing the metric of
AdS$_{d+1}$ in global coordinates
\[
-\cosh^{2}\rho~d\tau^{2}+d\rho^{2}+\sinh^{2}\rho~\left(  d\theta^{2}+\sin
^{2}\theta~d\Omega_{d-2}^{2}\right)  ~,
\]
we can analyze the geodesic motion of a massless particle of energy $E/2$ and spin $J/2$, conjugate to translations in $\tau$ and $\theta$. This gives the first order equation
\[
\dot{\rho}^{2}+\frac{J^{2}}{4\sinh^{2}\rho}=\frac{E^{2}}{4\cosh^{2}\rho}
\]
where $E=2\cosh^{2}\left(  \rho\right)  \dot{\tau}$ and
$J=2\sinh^{2}\left( \rho\right) \dot{\theta}$, and where the dot
denotes differentiation with respect to an affine parameter. The
particle reaches a minimum geodesic distance
$\rho_{\mathrm{\min}}$ to the origin when $\dot{\rho}=0$, that is
at $\tanh\rho_{\min}=J/E$. We now consider two particles in a
symmetric collision with total energy $E$ and spin $J$, as in
figure \ref{fig5}.
They reach a minimum relative geodesic distance $r=2\rho_{\min}$ given by
\[
\tanh\left(  \frac{r}{2}\right)  =\frac{J}{E}~,
\]
thus justifying geometrically the definition (\ref{ImpactParam}).


\subsection{Free Propagation\label{SectFree}}

\begin{figure}
[ptb]
\begin{center}
\includegraphics[height=1.737in,width=4.1766in]
{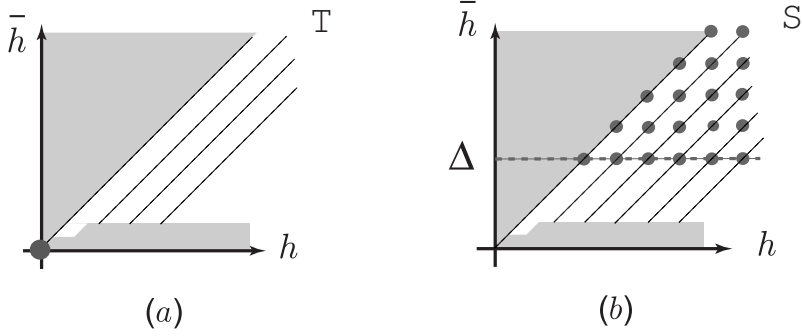} \caption{Partial wave decomposition of the amplitude
$\mathcal{A}_{0}=1$ corresponding to free propagation in AdS in
figure $1(a)$. In the \emph{T}--channel decomposition
only the vacuum contributes, whereas the \emph{S}--channel
decomposition receives contributions from a full lattice of
$\mathcal{O}_{1}\mathcal{\partial}^{N}\mathcal{O}_{2}$ composites
of dimensions $h,\bar {h}\in\Delta+\mathbb{N}_{0}$.} \label{fig6}
\end{center}
\end{figure}

The four--point amplitude $\mathcal{A}$ can be described dually as
gravitational interaction in AdS space. In particular, neglecting
string corrections, we have the expansion
\[
\mathcal{A}=\mathcal{A}_{0}+\mathcal{A}_{1}+\mathcal{A}_{2}+\cdots,
\]
in powers of the coupling constant $G$. The above expansion starts with the
contribution from the disconnected graph in figure \ref{fig4}$(a)$, which
describes free propagation in AdS and is given by the product of two--point
functions $\left\langle \mathcal{O}_{1}\left(  \mathbf{p}_{1}\right)
\mathcal{O}_{1}\left(  \mathbf{p}_{3}\right)  \right\rangle \left\langle
\mathcal{O}_{2}\left(  \mathbf{p}_{2}\right)  \mathcal{O}_{2}\left(
\mathbf{p}_{4}\right)  \right\rangle $. Choosing appropriate normalization for
the external operators $\mathcal{O}_{1}$, $\mathcal{O}_{2}$ we have that
\[
\mathcal{A}_{0}=1~.
\]
From the graph, it is intuitively clear that, in the \emph{T}--channel, only
the vacuum state with $h=\bar{h}=0$ contributes, as shown in figure
\ref{fig6}$(a)$.
In fact, with an appropriate normalization for
$\mathcal{T}_{h,\bar{h}}$ we have that
\begin{equation}
\mathcal{A}_{0}=\mathcal{T}_{0,0}~.\label{tfree}
\end{equation}

\noindent
On the other hand, the \emph{S}--channel decomposition of $\mathcal{A}_{0}$ is
more subtle. In fact, as shown in figure \ref{fig6}$(b)$, we expect that the
composites
\[
\mathcal{O}_{1}\partial_{\mu_{1}}\cdots\partial_{\mu_{J}}
\partial^{2n} \mathcal{O}_{2}\, ,
\]
of dimension $E=2\Delta+J+2n$ and spin $J$, contribute to the
\emph{S}--channel decomposition and define a lattice of operators of dimension
$h=\Delta+J+n$, $\bar{h}=\Delta+n$ given by
\[
h,\bar{h}\in\Delta+\mathbb{N}_{0}\text{ }
,~\ \ \ \ \ \ \ \ \ \ \ \ \ \ \ \ \ \ \ \Delta\leq\bar{h}\leq h~.
\]
Again, with an appropriate normalization for the \emph{S}--channel
partial waves $\mathcal{S}_{h,\bar{h}}$, we have the decomposition
\begin{equation}
\mathcal{A}_{0}=\sum_{\Delta\leq\bar{h}\leq h}\ \mathcal{S}_{h,\bar{h}
}~,\label{sfree}
\end{equation}

\noindent where it is understood that the sum is restricted to
$h,\bar{h}\in \Delta+\mathbb{N}_{0}$. Finally, we have the impact
parameter representation corresponding to (\ref{sfree})
\[
\mathcal{A}_{0}=\int_{0}^{\infty}dh\int_{0}^{h}d\bar{h}~~~\mathcal{I}
_{h,\bar{h}}~.
\]
Note that the normalizations chosen for the specific case $d=2$ are compatible
with (\ref{tfree}) and (\ref{sfree}).


\subsection{Tree--Level Interaction in the \emph{S}--Channel\label{TreeS}}


We expect that the full amplitude $\mathcal{A}$ can be expanded in
\emph{S}--channel partial waves as
\begin{equation*}
\mathcal{A}\simeq\sum_{h\geq\bar{h}\geq\Delta}\left(  1+R(h,\bar{h})\right)
~\mathcal{S}_{h+\Gamma(h,\bar{h}),~\bar{h}+\Gamma(h,\bar{h})}
~\ ,
\end{equation*}
\noindent where $2\Gamma(h,\bar{h})$ is the anomalous dimension of
the intermediate primary and $R(h,\bar{h})$ is related to its
three--point coupling to the external operators.  We are assuming,
by analogy with the flat space situation, that in the large $h$,
$\bar{h}$ limit the relevant intermediate primaries are the
composites $\mathcal{O}_{1}\partial^{N} \mathcal{O}_{2} $ that
already contribute at leading order.  Let us now analyze the
tree--level graph $\mathcal{A}_{1}$ in figure \ref{fig2}$(a)$,
corresponding to the exchange of a spin $j$ massless particle in
\textrm{AdS}$_{d+1}$.  Denoting by $\sigma_{h,\bar{h}}$ and
$\rho_{h,\bar{h}}$ the contribution of this graph to
$\Gamma(h,\bar{h})$ and $R(h,\bar{h})$, we can write
\begin{equation}
\mathcal{A}_{1}=\sum_{\Delta\leq\bar{h}\leq h}\ \sigma_{h,\bar{h}}\left(
\frac{\partial}{\partial h}+\frac{\partial}{\partial\bar{h}}\right)
\mathcal{S}_{h,\bar{h}}+\sum_{\Delta\leq\bar{h}\leq h}\ \rho_{h,\bar{h}
}~\mathcal{S}_{h,\bar{h}}\ . \label{SchExp}
\end{equation}

\begin{figure}
[ptb]
\begin{center}
\includegraphics[height=1.2256in,width=2.176in]
{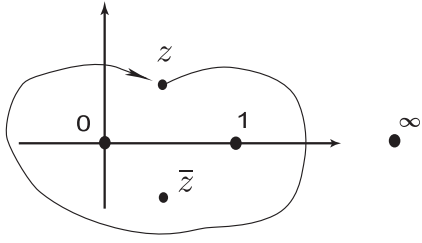}
\caption{Analytic continuation of $\mathcal{A}_{1}$ to obtain $\mathcal{A}_{1}^{\circlearrowright}$. The variable $\bar{z}$ is kept fixed and $z$ is
transported clockwise around the point at infinity, circling the points
$0,1,\bar{z}$.}
\label{fig9}
\end{center}
\end{figure}

The tree--level eikonal computation of the graph \ref{fig2}$(a)$
does not give an approximation to the amplitude $\mathcal{A}_{1}$,
as one would expect by analogy with the flat space result. Rather,
as was shown in \cite{ourSW}, it computes the small $z,\bar{z}$
behavior of the discontinuity function (or monodromy of the
amplitude around the point at infinity)
\[
\mathcal{M}_{1}\left(  z,\bar{z}\right) =\operatorname{Disc}{}_{z}
~\mathcal{A}_{1}\left( z,\bar{z}\right) \equiv\frac{1}{2\pi
i}\left( \mathcal{A}_{1}^{\circlearrowright}\left(
z,\bar{z}\right) -\mathcal{A} _{1}\left(  z,\bar{z}\right) \right)
,
\]
where $\mathcal{A}_{1}^{\circlearrowright}$ is the analytic continuation of
$\mathcal{A}_{1}$ obtained by keeping $\bar{z}$ fixed and by transporting $z$
clockwise around the point at infinity as in figure \ref{fig9}.
Since $\mathcal{S}_{h,\bar{h}}$ behaves around $z,\bar{z}\sim\infty$ as
$$
z^{\Delta-h}\bar{z}^{\Delta-\bar{h}}
\sum_{n,\bar{n}\geq0}
z^{-n}\bar{z}^{-\bar{n}}c_{n,\bar{n}}\left(  h,\bar{h}\right)  +\left(
z\leftrightarrow\bar{z}\right)  ~,
$$
we have that, for $h,\bar{h}\in\Delta+\mathbb{N}_{0}$,
\[
\left(  \frac{\partial}{\partial h}+\frac{\partial}{\partial\bar{h}}\right)
\mathcal{S}_{h,\bar{h}}=-\ln\left(  z\bar{z}\right)  \mathcal{S}_{h,\bar{h}
}+\cdots~,
\]
where the terms in the dots have an expansion around $z,\bar{z}\sim\infty$
with integer powers of $z^{-1},\bar{z}^{-1}$ and do not contribute to
$\mathcal{M}_{1}$. It is then clear that

\begin{equation}
\mathcal{M}_{1}\left(  z,\bar{z}\right)=\sum_{\Delta\leq\bar{h}\leq h}\ \sigma_{h,\bar{h}}
~\mathcal{S}_{h,\bar{h}}\left(  z,\bar{z}\right)\label{mono}
\end{equation}

\noindent
carries all the information about the anomalous dimensions $\sigma_{h,\bar{h}}$.

In \cite{ourSW} it was shown that the leading behavior of
$\mathcal{M}_{1}$ for small $z,\bar{z}$ is given by
\begin{align}
\mathcal{M}_{1}   \simeq &-8G\mathcal{N}_{\Delta_{1}}\mathcal{N}_{\Delta_{2}
}\Gamma\left(  2\Delta_{1}-1+j\right)  \Gamma\left(  2\Delta_{2}-1+j\right)
\times\label{eq300}\\
& \times~\left(  -q^{2}\right)  ^{\Delta_{1}}\left(  -p^{2}\right)
^{\Delta_{2}}\int_{\mathrm{H}_{d-1}}\widetilde{dx}\widetilde{dy}~\frac
{\Pi\left(  x,y\right)  }{\left(  2q\cdot x\right)
^{2\Delta_{1}-1+j}\left( 2p\cdot y\right)
^{2\Delta_{2}-1+j}}~,\nonumber
\end{align}
\noindent where $\Pi\left(  x,y\right)  $ is the Euclidean scalar
propagator on the hyperbolic space $\mathrm{H}_{d-1}$ with mass
squared $d-1$ given in (\ref{explicitprop}). The leading behavior
is then of the form
\[
\mathcal{M}_{1}\left(  z,\bar{z}\right)  \simeq\left(  -z\right)
^{1-j}M\left(  \frac{\bar{z}}{z}\right)  ~,
\]
with $M\left(  w\right)^{\star} = M \left( w^{\star}\right)$ and
$M\left( w\right) =w^{1-j}M\left( 1/w\right)$. Recall from section
\ref{impactsection} that, in the limit of small $z,\bar{z}$, we
may approximate $\mathcal{M}_{1}$ with the impact parameter
representation
\begin{equation}
\mathcal{M}_{1}\simeq\int dhd\bar{h}\ \sigma_{h,\bar{h}}~\mathcal{I}
_{h,\bar{h}}~,\label{eq200}
\end{equation}

\noindent where the leading behavior of $\sigma_{h,\bar{h}}$ for
large $h,\bar{h}$ determines the leading behavior of
$\mathcal{M}_{1}$ at small $z,\bar{z}$. To match (\ref{eq200})
with (\ref{eq300}) using the integral representation
(\ref{intrep}) it is convenient to express the integral over the
hyperboloids in (\ref{eq300}) as an integral over Milne wedges

\begin{align*}
\mathcal{M}_{1}  \simeq &-8G\mathcal{N}_{\Delta_{1}}\mathcal{N}_{\Delta_{2}
}\left(  -q^{2}\right)  ^{\Delta_{1}}\left(  -p^{2}\right)^{\Delta_{2}}
\times\\
& \times~\int_{M} \,\frac{d(t_1\, x) }{t_1^{d-2\Delta_{1}-j+1}}
\,\frac{d(t_2\, y) }{t_2^{d-2\Delta_{2}-j+1}}
\,e^{-2 q\cdot (t_1\, x) -2 p\cdot (t_2\,y)} \,\Pi\left(  x,y\right) \ .
\end{align*}

\noindent We then conclude that the leading behavior of the
anomalous dimensions $\sigma_{h,\bar{h}}$ for large $h,\bar{h}$ is
given by

\begin{equation}
\sigma_{h,\bar{h}}\simeq-8G~\left(  \frac{s}{4}\right)  ^{j-1}\Pi\left(
x,y\right)  ~,~\;\ \ \ \ \ \ \ \ \ \ \left(  s\rightarrow\infty\right),
\label{anomdim}
\end{equation}

\noindent where $s=4h\bar{h}$. The propagator $\Pi\left(
x,y\right)  $ depends only on the geodesic distance $r$ between
$x,y\in\mathrm{H}_{d-1}$, which is given by $\cosh r=-x\cdot y$
and it is related to the conformal dimensions by
\[
e^{-r}=\frac{\bar{h}}{h}~.
\]
Using the explicit form (\ref{explicitprop}) of the propagator, we\ can express
$\Pi$ in terms of the conformal dimensions $h,\bar{h}$ for various dimensions
$d$, as shown in the following table:
\[
\begin{tabular}
[c]{|c|c|}\hline
$d$ & $\Pi \spaup{.4}   \spadown{.2}$      \\\hline
$2$ & $\frac{1}{2}\frac{\bar{h}}{h} \spaup{.4}   \spadown{.2}      $\\\hline
$3$ & $-\frac{1}{2\pi}\left[  \frac{h^{2}+\bar{h}^{2}}{2h\bar{h}}\ln\left(
\frac{h-\bar{h}}{h+\bar{h}}\right)  +1\right]  \spaup{.5}   \spadown{.2} $\\\hline
$4$ & $\frac{1}{2\pi}\frac{\bar{h}^{3}}{h\left(  h^{2}-\bar{h}^{2}\right)  } \spaup{.4}   \spadown{.3}
$\\\hline
$5$ & $\frac{1}{8\pi^{2}}\left[  8\frac{h^{2}\bar{h}^{2}}{\left(  h^{2}
-\bar{h}^{2}\right)  ^{2}}+3\frac{h^{2}+\bar{h}^{2}}{h\bar{h}}\ln\left(
\frac{h-\bar{h}}{h+\bar{h}}\right)  +6\right]  \spaup{.6}   \spadown{.4} $\\\hline
$6$ & $\frac{1}{\pi^{2}}\frac{\bar{h}^{5}\left(  2h^{2}-\bar{h}^{2}\right)
}{h\left(  h^{2}-\bar{h}^{2}\right)  ^{3}} \spaup{.6}   \spadown{.3}$\\\hline
\end{tabular}
\]

The eikonal result (\ref{eq300}) allowed us to determine, in the
\emph{S}--channel partial wave decomposition (\ref{SchExp}), the
high spin and energy behavior of the anomalous dimensions
$\sigma_{h,\bar{h}}$, but not of the coefficients
$\rho_{h,\bar{h}}$. We will now show that in fact
\begin{equation}\rho_{h,\bar{h}}\simeq\partial \sigma_{h,\bar{h}}\label{boldone2}\end{equation} up to
sub--leading terms, where we use the notation
\[\partial=\partial_{h}+\partial_{\bar{h}}\, .\]
In this case, the high $h,\bar{h}$ behavior of the amplitude
$\mathcal{A}_{1}$ can be fully reconstructed from
$\mathcal{M}_{1}$ and is given by

\begin{equation}
\mathcal{A}_{1}\simeq\sum_{h\geq\bar{h}\geq\Delta}\partial\left(
\sigma_{h,\bar{h}}\mathcal{S}_{h,\bar{h}}\right)  ~.\label{boldone}
\end{equation}

\noindent
In order to show this, consider the part of $\mathcal{A}_{1}$ which is of the form $\sum\sigma_{h,\bar{h}}\partial \mathcal{S}_{h,\bar{h}}$.
It is approximated by $\int dhd\bar{h}~\left(
\partial\sigma_{h,\bar{h}}\right)  \mathcal{I}_{h,\bar{h}}$, where we have
replaced $\sum\rightarrow\int$,
$\mathcal{S}\rightarrow\mathcal{I}$ and where we have integrated
by parts. The leading behavior of the above integral for
$z,\bar{z}\rightarrow0$ is $\left(  z\bar{z}\right)
^{\frac{3}{4}-\frac{j} {2}}$, since
$\partial\sigma_{h,\bar{h}}\sim\left(  h\bar{h}\right)
^{j-\frac{3}{2}}$. On the other hand, the OPE expansion of
$\mathcal{A}_{1}$ is dominated by the exchange of the massless
particle and must start with $\left(  z\bar{z}\right)
^{\frac{d}{2}}$. Therefore, the term in $\mathcal{A}_{1}$ of the
form $\sum\rho_{h,\bar{h}}\mathcal{S}_{h,\bar{h}}$ must compensate
the anomalous dimension contribution, giving (\ref{boldone}),
which is a total derivative in the impact parameter approximation
and which is therefore sub--leading near $z,\bar{z}\sim0$.


\subsection{Tree--Level Interaction in the \emph{T}--Channel\label{TreeT}}


The function $\mathcal{M}_{1}$ can also be used to obtain
information regarding the decomposition of the tree--level graph
$\mathcal{A}_{1}$ in \emph{T}--channel partial waves. Decomposing
the amplitude $\mathcal{A}_{1}$ as

\begin{equation}
\mathcal{A}_{1}=\sum\ \mu_{h,\bar{h}}~\mathcal{T}_{h,\bar{h}}~,\label{AinT}
\end{equation}

\noindent
the function $\mathcal{M}_{1}$ can be written as
\[
\mathcal{M}_{1}=\sum\ \mu_{h,\bar{h}}\operatorname{Disc}{}_{z}~\mathcal{T}
_{h,\bar{h}}~.
\]
Thus, we must first analyze in detail the behavior of the functions
$\mathcal{T}_{h,\bar{h}}\left(  z,\bar{z}\right)  $ as we rotate the point $z$
clockwise around infinity, keeping $\bar{z}$ fixed. Since the
eikonal result (\ref{eq300}) holds around $z,\bar{z}\sim0$, we shall need only
the leading behavior of $\operatorname{Disc}{}_{z}~\mathcal{T}_{h,\bar{h}}$
around the origin.

Consider first the behavior of $\mathcal{T}_{h,\bar{h}}$ in the limit $\bar
{z}\rightarrow0$, with $z$ fix. The operator $D_{T}$ in (\ref{Tchannel}) reduces to
\[
z^{2}\left(  1-z\right)  \partial^{2}-z^{2}\partial+\bar{z}^{2}\bar{\partial
}^{2}-\left(  d-2\right)  \bar{z}\bar{\partial}~.
\]
Using the boundary condition $\mathcal{T}_{h,\bar{h}}\sim\left(  -z\right)
^{h}\left(  -\bar{z}\right)  ^{\bar{h}}$ around the origin, we conclude that
\[
\mathcal{T}_{h,\bar{h}}\sim\left(  -\bar{z}\right)  ^{\bar{h}}\left(
-z\right)  ^{h}F\Big(  h,h,2h\Big| z\Big)  ~.
\]
Since
\begin{align*}
\left(  -z\right)  ^{h}F\Big(  h,h,2h \Big| z\Big)   & =\frac{\Gamma\left(
2h\right)  }{\Gamma\left(  h\right)  ^{2}}\sum_{n\geq0}\frac{\left(  h\right)
_{n}\left(  1-h\right)  _{n}}{\left(  n!\right)  ^{2}}z^{-n}\times\\
& \times\bigg(  \ln\left(  -z\right)  +2\psi\left(  n+1\right)  -\psi\left(
h+n\right)  -\psi\left(  h-n\right)  \bigg)
\end{align*}
we obtain that
\[
\operatorname{Disc}{}_{z}~\mathcal{T}_{h,\bar{h}}\sim-\frac{\Gamma\left(
2h\right)  }{\Gamma\left(  h\right)  ^{2}}\left(  -\bar{z}\right)
^{\bar{h} }F\Big(  h,1-h,1 \Big| z^{-1}\Big) ~.
\]
In the limit of small $z$ the leading behavior is

\begin{equation}
\operatorname{Disc}{}_{z}~\mathcal{T}_{h,\bar{h}}\sim-\frac{\Gamma\left(
2h\right)  \Gamma\left(  2h-1\right)  }{\Gamma\left(  h\right)  ^{4}}~\left(
-z\right)  ^{1-h}\left(  -\bar{z}\right)  ^{\bar{h}}~.
\label{bcDiscT}
\end{equation}

\noindent
Recall that we derived this result in the limit $\bar{z}\rightarrow0$. To
understand the general behavior around $z,\bar{z}\sim0$ of
$\operatorname{Disc}{}_{z}~\mathcal{T}_{h,\bar{h}}$, we expand $\mathcal{T}
_{h,\bar{h}}$ in powers of $\bar{z}$ as

\begin{equation}
\mathcal{T}_{h,\bar{h}}\sim\sum_{n\geq0}\left(  -\bar{z}\right)  ^{\bar{h}
+n}g_{n}\left(  z\right)  ~.\label{eqTex}
\end{equation}

\noindent
We have just determined that $\operatorname{Disc}{}_{z}~g_{0}\sim z^{1-h}$ for
small $z$. The other functions $g_{n}$ are determined recursively by expanding
the differential equation $D_{T}=c_{h,\bar{h}}$ in powers of $\bar{z}$. A
rather cumbersome but straightforward computation shows that
$\operatorname{Disc}{}_{z}~g_{n}\sim z^{1-h-n}$ for small $z$. Therefore, we
conclude in general that

\begin{equation}
\operatorname{Disc}{}_{z}~\mathcal{T}_{h,\bar{h}}\sim\left( -z\right)
^{1-h}\left(  -\bar{z}\right)  ^{\bar{h}}~G_{h,\bar{h}}\left( \frac{\bar{z}
}{z}\right) \label{discT}
\end{equation}

\noindent
around $z,\bar{z}\sim0$. The function $G_{h,\bar{h}}\left(  w\right)  $ is
regular around $w=0$ and, using (\ref{bcDiscT}), satisfies $G_{h,\bar{h}}\left(  0\right) =-\Gamma\left(  2h\right)  \Gamma\left(  2h-1\right)  /\Gamma\left(  h\right)^{4}$. The careful reader will have noticed that in equation (\ref{eqTex}) we have implicitly neglected to symmetrize in $z\leftrightarrow\bar{z}$. Had we not, the function $\mathcal{T}_{h,\bar{h}}$ would have had an extra
contribution of the form $\sum_{n\geq0}\left(  -\bar{z}\right)  ^{h+n}
f_{n}\left(  z\right)  $ with $\operatorname{Disc}{}_{z}~f_{n}\sim
z^{1-\bar{h}-n}$. These terms then give sub--leading contributions to
(\ref{discT}).

To compute explicitly the function $G_{h,\bar{h}}$, we consider the operator
$D_{T}$ in (\ref{Tchannel}) near $z,\bar{z}\sim0$, which reduces to
\[
z^{2}\partial^{2}+\bar{z}^{2}\bar{\partial}^{2}+\left(  d-2\right)
\frac{z\bar{z}}{z-\bar{z}}\left(  \partial-\bar{\partial}\right)  ~.
\]
Acting on (\ref{discT}) the differential equation $D_{T}=c_{h,\bar{h}}$
becomes of the hypergeometric form
\[
2w\left(  1-w\right)  G^{\prime\prime}+\left[  2\left(  h+\bar{h}\right)
\left(  1-w\right)  -\left(  d-2\right)  \left(  1+w\right)  \right]
G^{\prime}=\left(  d-2\right)  \left(  h+\bar{h}-1\right)  G~,
\]
in terms of $w=\bar{z}/z$. We then arrive at the result

\begin{equation}
G_{h,\bar{h}}=-\frac{\Gamma\left(  2h\right)  \Gamma\left(  2h-1\right)
}{\Gamma\left(  h\right)  ^{4}}~F \left. \left(  \frac{d}{2}-1,h+\bar{h}-1,h+\bar {h}+1-\frac{d}{2} \right| \frac{\bar{z}}{z}\right)  ~. \label{Ghhbar}
\end{equation}

We are now in a position to determine the implications of the eikonal result
(\ref{eq300}) for the \emph{T}--channel expansion coefficients $\mu_{h,\bar
{h}}$ in (\ref{AinT}).
\begin{figure}
[ptb]
\begin{center}
\includegraphics[height=1.8853in,width=2.7032in]
{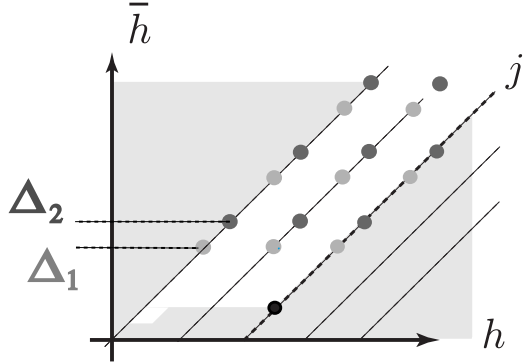} \caption{\emph{T}--channel decomposition of the spin
$j$ exchange graph in figure $2(a)$. Only partial waves
with spin $J\leq j$ contribute. The eikonal integral determines
the contributions with $J=j$, along the thick dashed line. We
display with a black dot the contribution from the operator dual
to the exchanged particle, with spin $j$ and energy $d-j$. We also
show, with light and dark grey dots, the contributions from the
$\mathcal{O} _{1}\partial^{N}\mathcal{O}_{1}$ and
$\mathcal{O}_{2}\partial^{N} \mathcal{O}_{2}$ composites, with
dimensions $h,\bar{h}\in\Delta _{1}+\mathbb{N}_{0}$ and
$h,\bar{h}\in\Delta_{2}+\mathbb{N}_{0}$, respectively.}
\label{fig7}
\end{center}
\end{figure}
Recall that the leading behavior of $\mathcal{M}_{1}$ for small $z,\bar{z}$ is
controlled by
\[
\mathcal{M}_{1}\simeq\left( -z\right)^{1-j}M\left( \frac{\bar{z}}{z}\right),
\]
as derived in the dual \textrm{AdS} description. We therefore
immediately conclude from (\ref{discT}) that, in the decomposition
(\ref{AinT}), only partial waves with
\[
J=h-\bar{h}\leq j
\]
can appear, as shown in figure \ref{fig7}. Moreover, the coefficients
$\mu_{h,\bar{h}}$ for $J=h-\bar{h}=j$ are determined directly by the function
$M\left( w\right)$ from

\begin{equation}
\sum_{h-\bar{h}=j}\mu_{h,\bar{h}}~w^{\bar{h}}~G_{h,\bar{h}}\left(  w\right)
=M\left(  w\right)  ~.\label{mucoef}
\end{equation}

\noindent
In the simple case of $d=2$ the functions $G_{h,\bar{h}}$ are constants, so
that the coefficients $\mu_{h,\bar{h}}$ are simply obtained by expanding $M$
in increasing powers of $w$.


\subsection{An Example in $d=2$\label{SectExample}}


To be more explicit we conclude this section with a simple example where we
can check our results. We shall consider the case $d=\Delta_2=2$
and $j=0$ corresponding to massless scalar exchange in AdS$_3$.
As shown in \cite{ourSW}, the basic amplitude $\mathcal{A}_{1}$ is given by

\begin{equation}
\mathcal{A}_{1}=\frac{8G}{\pi}~\mathbf{p}_{13}^{\Delta_{1}}\mathbf{p}
_{24}~D_{\Delta_{1},\Delta_{1},1,1}^{2}\left(  \mathbf{p}_{1},\mathbf{p}
_{3},\mathbf{p}_{2},\mathbf{p}_{4}\right)  ~.\label{ampex}
\end{equation}

\noindent where $D^d_{\Delta_i}\left(  \mathbf{p}_{i} \right)$ are
the standard $D$--functions \cite{FreedmanRev, ourSW}. On the
other hand, in \cite{ourSW} we have also explicitly computed the
integral (\ref{eq300}), which controls the leading behavior of
$\mathcal{M}_{1}=\operatorname*{Disc}_{z}\mathcal{A}_{1}$ as
$z,\bar{z}\rightarrow0$, obtaining $
\mathcal{M}_{1}\simeq-zM\left(\bar{z}/{z}\right),$ with $M\left(
w\right)$ given by

\begin{equation}
M\left(  w\right)  =-\frac{4G}{2\Delta_{1}-1}\,w F \Big(  1,\Delta_{1}
,2\Delta_{1}\Big|1-w\Big)  . \label{exampleM}
\end{equation}

\begin{figure}
[ptb]
\begin{center}
\includegraphics[height=0.5815in,width=4.8487in]
{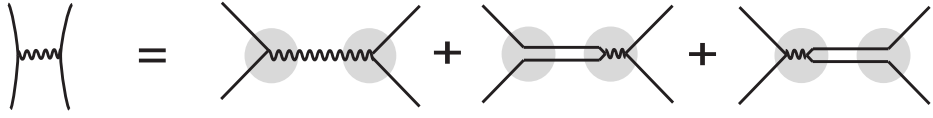}
\caption{\emph{T}--channel decomposition of the graph in figure $2(a)$.
The contributions come from the operator dual to the exchanged particle as well
as from the $\mathcal{O}_{1}\partial^{N}\mathcal{O}_{1}$ and $\mathcal{O}_{2}\partial^{N}\mathcal{O}_{2}$ composites.}
\label{fig8}
\end{center}
\end{figure}

Recall, from the discussion in section \ref{TreeT}, that the function
$M\left(  w\right)  $ contains information about the \emph{T}--channel
expansion of $\mathcal{A}_{1}$
\[
\mathcal{A}_{1}=\sum_{h}\mu_{h,h}~\mathcal{T}_{h,h}~,
\]
which has only spin zero contributions. More precisely, the expansion (\ref{mucoef}), with the explicit form of $G_{h,\bar{h}}$ in (\ref{Ghhbar}), becomes
\[
M\left(  w\right)  =-\sum_{h}\mu_{h,h}~\frac{\Gamma\left(  2h\right)
\Gamma\left(  2h-1\right)  }{\Gamma\left(  h\right)  ^{4}}w^{h}~.
\]
Using the standard properties of the hypergeometric functions, we
may expand $M\left(  w\right)  $ in (\ref{exampleM}) around $w=0$
as
\begin{align*}
M\left(  w\right)   & =4G\frac{\Gamma\left(  1-\Delta_{1}\right)  }
{\Gamma\left(  \Delta_{1}\right)  }\sum_{h\in1+\mathbb{N}_{0}}\frac
{\Gamma\left(  \Delta_{1}-1+h\right)  }{\Gamma\left(  1-\Delta_{1}+h\right)
}w^{h}-\\
& -4G\frac{\Gamma\left(  1-\Delta_{1}\right)  }{\Gamma\left(  \Delta
_{1}\right)  }\sum_{h\in\Delta_{1}+\mathbb{N}_{0}}\frac{\Gamma\left(
\Delta_{1}-1+h\right)  }{\Gamma\left(  1-\Delta_{1}+h\right)  }w^{h},
\end{align*}

\noindent
finally concluding that
\[
\mathcal{A}_{1}=-\mu_{1}\mathcal{T}_{1,1}+\sum_{h\in\Delta_{1}+\mathbb{N}_{0}
}\mu_{h}\mathcal{T}_{h,h}-\sum_{h\in\Delta_2+\mathbb{N}_{0}}\mu_{h}\mathcal{T}_{h,h}~,
\]
with
\[
\mu_{h}=4G\frac{\Gamma\left(  1-\Delta_{1}\right)  }{\Gamma\left(  \Delta
_{1}\right)  }\frac{\Gamma\left(  \Delta_{1}-1+h\right)  }{\Gamma\left(
1-\Delta_{1}+h\right)  }\frac{\Gamma\left(  h\right)  ^{4}}{\Gamma\left(
2h\right)  \Gamma\left(  2h-1\right)  }~.
\]
The contributions come from the operator dual to the exchanged
particle, with spin $j=0$ and energy $2$, as well as from the
$\mathcal{O}_{1}
\partial^{N}\mathcal{O}_{1}$ and $\mathcal{O}_{2}\partial^{N}\mathcal{O}_{2}$
composites, with dimensions $h=\bar{h}\in\Delta_{1}+\mathbb{N}_{0}$ and
$h=\bar{h}\in\Delta_{2}+\mathbb{N}_{0}$ respectively, as shown in figures
\ref{fig7} and \ref{fig8}.

Finally let us consider the expansion of $\mathcal{A}_{1}\,$\ in the
$S$--channel. We shall restrict further our attention to the special case of
$\Delta_{1}=2$, where the amplitude (\ref{ampex}) can be explicitly computed
\cite{Bianchi}
\[
\mathcal{A}_{1}\left(  z,\bar{z}\right)  =-8G\frac{z^{2}\bar{z}^{2}}{\left(
\bar{z}-z\right)  }\left[  \frac{1}{1-\bar{z}}\partial-\frac{1}{1-z}
\bar{\partial}\right]  a\left(  z,\bar{z}\right)  ~,
\]
where
\[
a\left(  z,\bar{z}\right)  =\frac{\left(  1-z\right)  \left(  1-\bar
{z}\right)  }{\left(  z-\bar{z}\right)  }\left[  \mathrm{Li}_{2}\left(
z\right)  -\mathrm{Li}_{2}\left(  \bar{z}\right)  +\frac{1}{2}\mathrm{\ln
}\left(  z\bar{z}\right)  \mathrm{\ln}\left(  \frac{1-z}{1-\bar{z}}\right)
\right]  ~,
\]
with $\mathrm{Li}_{2}$ the dilogarithm. Using a symbolic
manipulation program we can check the decomposition (\ref{SchExp})
up to very high order, with
\begin{align*}
\sigma_{h,\bar{h}}  & =-\frac{4G}{h\left(  h-1\right)  }~,\\
\rho_{h,\bar{h}}  & =\left(  \frac{\partial}{\partial h}+\frac{\partial
}{\partial\bar{h}}\right)  \sigma_{h,\bar{h}}+\frac{4G}{h\left(  h-1\right)
\bar{h}\left(  \bar{h}-1\right)  }~.
\end{align*}
\noindent In the limit of large dimensions \thinspace$h$,$\bar{h}
\rightarrow\infty$, we have $\sigma_{h,\bar{h}}\sim-4G/h^{2}$, as
predicted from the general formula (\ref{anomdim}). In the same
limit, we have $\rho_{h,\bar{h}}\simeq\partial
\sigma_{h,\bar{h}}$, in agreement with (\ref{boldone2}).


\section{Graviton Dominance\label{GravDom}}


We have analyzed in great detail the tree--level exchange of a
spin $j$ particle in the \emph{T}--channel, given by graph
\ref{fig2}$(a)$ in the introduction. We have noticed that,
decomposing this graph in the \emph{S}--channel and considering
its contribution to partial waves of large spin and energy, the
dominant amplitude has the maximal value for the spin $j$ of the
exchanged particle. In gravitational theories in \textrm{AdS},
this particle is the graviton, whose exchange dominates the
interaction and determines the tree--level anomalous dimensions of
the double trace $\mathcal{O}_{1}\mathcal{O}_{2}$ composites to be
$2\sigma_{h,\bar{h}}\simeq -4Gs\Pi\left( x,y\right)$ for large
$h,\bar{h} $. On the other hand, the full gravitational theory in
\textrm{AdS} will have more interactions at tree--level, like $S$
and $U$--channel exchanges, as well as contact and non--minimal
interactions. Just as in flat space, though, all these other
interactions are subdominant in the large spin and energy limit,
and can be neglected in first approximation. We will not give a
complete proof of this fact, but we shall rather concentrate on
some specific significant examples. In particular, we will analyze
the graphs already considered in the introduction (in figure
\ref{fig2}), and we will concentrate on the case
$\Delta_{1}=\Delta_{2}$ for simplicity.

Letting $\mathcal{A}_{1}(z,\bar{z})$ be the amplitude for graph
\ref{fig2}$(a)$,  the amplitudes for graphs \ref{fig2}$(b)$ and
\ref{fig2}$(c)$ are simply obtained by permuting the external
particles and are given explicitly by
\begin{align*}
\left( z\bar{z}\right)^{\Delta}\mathcal{A}_{1}\left( \frac{1}{z}, \frac{1}{\bar{z}} \right),
\text{ \ \ \ \ \ \ \ \ \ \ \ \ \ \ \ \ \ \ \ \ \ \ \ \ \ }  & \text{(graph
}2(b)\text{)},\\
\left(  \frac{z\bar{z}}{\left(  1-z\right)  \left(
1-\bar{z}\right) }\right)  ^{\Delta}\mathcal{A}_{1}\left(
1-z,1-\bar{z}\right),  \text{ \ \ \ \ \ \ \ \ \ \ \ \ \ \ \ \ \ \
\ \ \ }  & \text{(graph }2(c)\text{)}.
\end{align*}

\noindent The $S$--channel decomposition of graph \ref{fig2}$(b)$
can be trivially deduced from the results of section
(\ref{TreeT}), which considers the mirror $T$--channel
decomposition of graph \ref{fig2}$(a)$. Without any further
analysis, we conclude that \ref{fig2}$(b)$ contributes only to
$S$--channel partial waves of spin $J\leq j$, and is therefore
local in spin as in flat space. It is therefore irrelevant when
considering large spin and energy decomposition of the full
tree--level amplitude.

To analyze the $S$--channel decomposition of graph \ref{fig2}$(c)$, let
us first note that the differential operator $D_{S}$ in (\ref{Schannel}) is
invariant under $z\rightarrow1-z$ and $\bar{z}\rightarrow1-\bar{z}$, whenever
$2\nu=\Delta_1-\Delta_2=0$. We therefore conclude that
\[
\left(  \frac{z\bar{z}}{\left(  1-z\right)  \left(  1-\bar{z}\right)
}\right)  ^{\Delta}\mathcal{S}_{h,\bar{h}}\left(  1-z,1-\bar{z}\right)
=\left(  -\right)  ^{h-\bar{h}}\mathcal{S}_{h,\bar{h}}\left(  z,\bar
{z}\right)  ~,
\]
where the normalization is fixed by recalling the leading behavior
of $\mathcal{S}_{h,\bar{h}}\sim
z^{\Delta-h}\bar{z}^{\Delta-\bar{h}}$ when
$z,\bar{z}\rightarrow\infty$. In fact, if we choose $z=i\lambda$,
$\bar {z}=-i\lambda$ to avoid branch cuts on the real axis,
 we have that $\mathcal{S}_{h,\bar{h}}\left(  z,\bar{z}\right)
\sim\left(  i\lambda\right)  ^{\Delta-h}\left(  -i\lambda\right)
^{\Delta-\bar{h}}$ and that $\mathcal{S}_{h,\bar{h}}\left(  1-z,1-\bar
{z}\right)  \sim\left(  -i\lambda\right)  ^{\Delta-h}\left(  i\lambda\right)
^{\Delta-\bar{h}}$, thus fixing the relative normalization to $\left(
-\right)  ^{h-\bar{h}}$. We then conclude that, if the $S$--channel expansion
of $\mathcal{A}_{1}$ is given by (\ref{SchExp}), then graph \ref{fig2}$(c)$ is given by
\[
\sum_{\Delta\leq\bar{h}\leq h}\ \left(  -\right)  ^{h-\bar{h}}~\sigma
_{h,\bar{h}}\left(  \frac{\partial}{\partial h}+\frac{\partial}{\partial
\bar{h}}\right)  \mathcal{S}_{h,\bar{h}}+\sum_{\Delta\leq\bar{h}\leq h}\left(
-\right)  ^{h-\bar{h}}\ \rho_{h,\bar{h}}~\mathcal{S}_{h,\bar{h}}~.
\]
Therefore, for instance, the extra contribution to the anomalous dimension is
given by
\[
\left(  -\right)  ^{h-\bar{h}}~2\sigma_{h,\bar{h}}~,
\]
which oscillates as a function of spin. In a coarse--grained
picture in which we consider large impact parameters and
continuous spins, these oscillations average to a vanishing
function, just as in the flat space case considered in the
introduction.

Finally, let us analyze the contact interaction of graph
\ref{fig2}$(d)$. Using the techniques developed in \cite{ourSW},
one can easily establish that the discontinuity
$\operatorname*{Disc} {}_{z}$ of the amplitude corresponding to
graph \ref{fig2}$(d)$ is proportional to
\begin{equation}
\left(  -q^{2}\right)  ^{\Delta}\left(  -p^{2}\right)  ^{\Delta}
\int_{\mathrm{H}_{d-1}}\widetilde{dx}~\frac{1}{\left(  2q\cdot
x\right) ^{2\Delta-1}\left(  2p\cdot x\right)
^{2\Delta-1}}~.\label{contactint}
\end{equation}

\noindent This expression is essentially equation (\ref{eq300})
with $j=0$ and the propagator $\Pi\left( x,y\right)$ replaced by
the delta function $\delta\left(  x,y\right)$ on the transverse
space $\mathrm{H}_{d-1}$. It is therefore easy to follow section
\ref{SecPropagation} and interpret it in the \emph{S} and in the
\emph{T}--channel partial wave expansions. On one hand, the
discontinuity function is an impact parameter approximation to the
high spin and energy anomalous dimensions, which are now
proportional to
\[
\frac{1}{s}\delta\left(  x,y\right)  ~,
\]
where we recall that $s=4h\bar{h}$ and $-2x\cdot
y=\bar{h}/h+h/\bar{h}$. Therefore the delta function fixes
$h=\bar{h}$, \textit{i.e.}, spin $J=0$. On the other hand, since
the discontinuity function (\ref{contactint}) goes like $z$ for
small $z,\bar{z}$, we have only spin $J=0$ partial waves appearing
in the \emph{T}--channel decomposition. Expression
(\ref{contactint}) also generates the relative weights of all
these spin zero contributions. In both channels, as expected,
graph \ref{fig2}$(d)$ only contains spin zero intermediate
primaries, and therefore does not effect the large spin results
which are only controlled by graph \ref{fig2}$(a)$.


\section{A Conjecture and Conclusions\label{Conj}}


To complete the eikonal program, one crucial step is missing. In
flat space, one can approximately reconstruct the full amplitude
from the phase shift $\sigma$ using (\ref{eikonalre--sum}). This
step cannot be immediately done in $\mathrm{AdS}_{d+1}$, even at
tree--level, since the eikonal two--point function computed in
\cite{ourSW} determines only the leading behavior of the
\textit{discontinuity} $\mathcal{M}_{1}$ of the relevant
tree--level amplitude $\mathcal{A}_{1}$ in figure \ref{fig2}a.
Nevertheless, we can be bold and try to reconstruct the full
amplitude $\mathcal{A}$ from $\sigma_{h,\bar{h}}$. We shall assume
that $\mathcal{A}$ is dominated at large $h,\bar{h}$ only by the
$\mathcal{O}_{1}\mathcal{O}_{2}$ composites with finite anomalous
dimensions $\Gamma(h,\bar{h})$. This implies a decomposition of
the form

\begin{equation}
\mathcal{A}\simeq\sum_{h\geq\bar{h}\geq\Delta}\left(  1+R(h,\bar{h})\right)
~\mathcal{S}_{h+\Gamma(h,\bar{h}),~\bar{h}+\Gamma(h,\bar{h})}
~\ ,\label{generalex}
\end{equation}

\noindent
where $R(h,\bar{h})$ are now finite coefficients. Expanding in powers of
$\Gamma$ and dropping the explicit reference to $h,\bar{h}$, the above
equation reads
\[
\mathcal{A}\simeq\sum~\left(  1+R\right)  \left(  1+\Gamma\partial+\frac{1}
{2}\Gamma^{2}\partial^{2}+\cdots\right)~\mathcal{S}\text{~}.
\]
The coefficients $R=R_{1}+R_{2}+\cdots$ and the anomalous dimensions
$\Gamma=\Gamma_{1}+\Gamma_{2}+\cdots$ are computed in perturbation theory,
with $\Gamma_{1}\simeq\sigma$ and $R_{1}\simeq\partial\sigma$. Order by order
in the loop expansion, we then have the following expansions

\begin{align*}
\mathcal{A}_{0}  & \simeq\sum~\mathcal{S}~,\\
\mathcal{A}_{1}  & \simeq\sum~R_{1}\,\mathcal{S}+\Gamma_{1}\,\partial
\mathcal{S}~,\\
\mathcal{A}_{2}  & \simeq\sum~R_{2}\,\mathcal{S}+\left(  \Gamma_{2}
+R_{1}\Gamma_{1}\right)  \,\partial\mathcal{S+}\frac{1}{2}\Gamma_{1}
^{2}\,\partial^{2}\mathcal{S}~,\\
&\cdots
\end{align*}

\noindent
Since $\Gamma_{1}\simeq\sigma$, it is tempting to assume, in analogy with
(\ref{boldone}), that in general
\[
\mathcal{A}_{n}\simeq\frac{1}{n!}\sum~\partial^{n}\,\left(  \sigma
^{n}\mathcal{S}\right)  ~.
\]
This is compatible with (\ref{generalex}) provided that
\[
\Gamma\simeq\sum_{n\geq1}\frac{1}{n!}\partial^{n-1}\sigma^{n}
~,~\ \ \ \ \ \ \ \ \ \ \ \ \ \ \ R=\partial\Gamma~.\,
\]
For example, consider graviton exchange in $d=2$, where $\sigma\simeq
-4G\bar{h}^{2}$. The sum over $n$ can be done explicitly, arriving at the
following intriguing formula for the anomalous dimension
\[
\Gamma\simeq\frac{2\bar{h}}{1+\sqrt{1+16G\bar{h}}}-\bar{h}
\ ,\ \ \ \ \ \ \ \ \ \ \ \ \ \ \ \ \ \ \ \ \ \ \ \ \ \ \ \ \left(
j=2,~d=2\right).
\]
Clearly the results above are purely conjectural, and we must leave a complete
discussion of these issues for future research.

Finally, it would be very interesting to apply the techniques
developed in this paper to specific realizations of the AdS/CFT
correspondence. All of our discussion has been carried out for
pure AdS, decomposing the amplitudes in conformal partial waves
related to the isometries $SO\left( 2,d\right)$ of the underlying
spacetime. The impact parameter representation has also been
derived accordingly. It is thus important to generalize the
discussion to AdS $\times$ compact spaces and to superspaces, with
partial waves and impact parameter representations appropriate for
the isometry group of the full (super)space, including internal
(and possibly super) symmetries. This could then allow us to test
our results against computations performed directly in CFT duals
at \textit{finite} $N$.

Even though all computations in this paper are done in the gravity
regime, neglecting string effects, it known \cite{ACV} that, in
some cases, string effects do not alter the general eikonal
results in flat space. It is therefore tempting to speculate that
some of the results of this paper could be already visible in
weakly coupled CFT's at finite $N$.

\section*{Acknowledgments}
Our research is supported in part by INFN, by the MIUR--COFIN contract 2003--023852, by the EU contracts MRTN--CT--2004--503369, MRTN--CT--2004--512194, by the INTAS contract 03--51--6346, by the NATO grant PST.CLG.978785 and by the FCT contract POCTI/FNU/38004/2001. LC is supported by the MIUR contract \textquotedblleft Rientro dei cervelli\textquotedblright \ part VII. MSC was partially supported by the FCT grant SFRH/BSAB/530/2005. JP is funded by the FCT fellowship SFRH/BD/9248/2002. \emph{Centro de F\'{\i}sica do Porto} is partially funded by FCT through the POCTI programme.

\vfill

\eject

\appendix


\section{Impact Parameter Representation}


In this appendix we show that the integral representation
(\ref{eq1001}) of the impact parameter representation partial wave
$\mathcal{I}_{h,\bar{h}}$ solves the differential equation
(\ref{PDEimpact}). We start by recalling the relevant kinematics.
Choosing the four external points $\mathbf{p}_{i}$ as
\begin{align*}
\mathbf{p}_{1}  & =\left(  0,1,0\right)
~,\ \ \ \ \ \ \ \ \ \ \ \ \ \ \ \ \ \ \ \ \ \ \ \mathbf{p}_{2}=-\left(
1,p^{2},p\right)  ~,\\
\mathbf{p}_{3}  & =-\left(  q^{2},1,q\right)
~,\ \ \ \ \ \ \ \ \ \ \ \ \ \ \ \ \ \ \ \mathbf{p}_{4}=\left(  1,0,0\right)
~,
\end{align*}
the cross ratios $z,\bar{z}$ are determined by
\[
z\bar{z}=q^{2}p^{2},~\ \ \ \ \ \ \ \ \ \ \ \ \ \ \ \ z+\bar{z}=2p\cdot q~.
\]
In what follows, we choose once and for all a fixed point
$p\in-\mathrm{H} _{d-1}$, so that $p^{2}=-1$. We then view the
\emph{S}--channel impact parameter amplitude
$\mathcal{I}_{h,\bar{h}}$ as a function just of $q$. Recall that,
in terms of $z,\bar{z}$, the function $\left(  z\bar{z}\right)
^{-\Delta}\mathcal{I}_{h,\bar{h}}$ satisfies the following
differential equation
\[
z\partial^{2}+\bar{z}\bar{\partial}^{2}+\partial+\bar{\partial}+\frac
{d-2}{z-\bar{z}}\left(  z\partial-\bar{z}\bar{\partial}\right)  =\frac{\nu
^{2}}{z}+\frac{\nu^{2}}{\bar{z}}-h^{2}-\bar{h}^{2}.
\]
A tedious computation shows that, in terms of $q$, the above equation can be
written as

\begin{equation}
\left(  q^{i}p^{j}-\frac{1}{2}\eta^{ij}q\cdot p\right)  \frac{\partial
}{\partial q^{i}}\frac{\partial}{\partial q^{j}}+\frac{d}{2}p^{i}
\frac{\partial}{\partial q^{i}}=\nu^{2}\frac{2p\cdot q}{q^{2}}+h^{2}+\bar
{h}^{2}~.\label{Aeq1}
\end{equation}

\noindent
Consider first the following function
\[
f\left(  x\right)  =\left\vert x\right\vert ^{d}\int_{\mathrm{M}
}dy~~e^{-2p\cdot y}~\delta\left(  2y\cdot x~+h^{2}+\bar{h}^{2}\right)
~\delta\left(  x^{2}y^{2}-h^{2}\bar{h}^{2}\right)  ~,
\]
where we integrate over the future Milne cone $\mathrm{M}\in\mathbb{M}^{d}$
given by $y^{2}\leq0,~y^{0}\geq0$. Changing integration variable to
\[
z=-x\frac{\left(  x\cdot y\right)  \left(  x\cdot p\right)  -x^{2}\left(
y\cdot p\right)  }{\left(  x\cdot p\right)  ^{2}+x^{2}}+px^{2}\frac{\left(
x\cdot p\right)  \left(  y\cdot p\right)  +\left(  y\cdot x\right)  }{\left(
x\cdot p\right)  ^{2}+x^{2}},
\]
with $z^{2}=-x^{2}y^{2}$, $z\cdot p=-y\cdot x$, $z\cdot x=-x^{2}p\cdot y$
and~$dz=\left\vert x\right\vert ^{d}dy$, we also have the integral
representation
\[
f\left(  x\right)  =\int_{\mathrm{M}}dz~~e^{-\frac{2x\cdot z}{x^{2}}}
~\delta\left(  2z\cdot p~-h^{2}-\bar{h}^{2}\right)  ~\delta\left(  z^{2}
+h^{2}\bar{h}^{2}\right)  ~,
\]
from which it is clear that the function $f\left( x\right)$ satisfies

\begin{equation}
~p^{i}\frac{\partial}{\partial\left(  x^{i}/x^{2}\right)  }f=\left(
x^{2}p^{j}-2p\cdot x~x^{j}\right)  \frac{\partial}{\partial x^{j}}f=-\left(
h^{2}+\bar{h}^{2}\right)  ~f~.\label{Aeq3}
\end{equation}

\noindent
We now consider the following function

\begin{equation}
g\left(  q\right)  =\left(  -q^{2}\right)  ^{\nu}\int_{\mathrm{M}}\frac
{dx}{\left\vert x\right\vert ^{d-4\nu}}~e^{-2q\cdot x}~f\left(  x\right)
~.\label{Aeq2}
\end{equation}

\noindent
We claim that $g\left( q\right)$ satisfies the differential equation
(\ref{Aeq1}). Replacing

\begin{align*}
\frac{\partial}{\partial q^{i}}  & \rightarrow2\left(  \nu\frac{q_{i}}{q^{2}
}-x_{i}\right), \\
\frac{\partial}{\partial q^{i}}\frac{\partial}{\partial q^{j}}  &
\rightarrow2\nu\left(  \frac{\eta_{ij}}{q^{2}}-2\frac{q_{i}q_{j}}{q^{4}
}\right)  +4\left(  \nu\frac{q_{i}}{q^{2}}-x_{i}\right)  \left(  \nu
\frac{q_{j}}{q^{2}}-x_{j}\right),
\end{align*}

\noindent
one can easily show that (\ref{Aeq1}) is equivalent to

\begin{align*}
& \left(  -q^{2}\right)  ^{\nu}\int_{\mathrm{M}}\frac{dx}{\left\vert
x\right\vert ^{d-4\nu}}~e^{-2q\cdot x}~\left[  4\left(  q\cdot x\right)
\left(  p\cdot x\right)  -2x^{2}\left(  q\cdot p\right)  -\right. \\
& \left.  -\left(  d+4\nu\right)  \left(  p\cdot x\right)  -h^{2}-\bar{h}
^{2}\right]  ~f\left(  x\right)  =0~.
\end{align*}

\noindent
Using (\ref{Aeq3}) the above is equivalent to

\begin{align*}
& \left(  -q^{2}\right)  ^{\nu}\int_{\mathrm{M}}\frac{dx}{\left\vert
x\right\vert ^{d-4\nu}}~e^{-2q\cdot x}~\left[  4\left(  q\cdot x\right)
\left(  p\cdot x\right)  -2x^{2}\left(  q\cdot p\right)  -\right. \\
& \left.  -\left(  d+4\nu\right)  \left(  p\cdot x\right)  +\left(  x^{2}
p^{j}-2p\cdot x~x^{j}\right)  \frac{\partial}{\partial x^{j}}\right]
~f\left(  x\right)  =0~,
\end{align*}

\noindent
which in turn is equal to
\[
\int_{\mathrm{M}}dx~\frac{\partial}{\partial x^{j}}\left[  \left(
-x^{2}\right)  ^{-\frac{d}{2}+2\nu}\left(  x^{2}p^{j}-2p\cdot x~x^{j}\right)
e^{-2q\cdot x}~f\left(  x\right)  \right]  =0~.
\]
This last equation is true since the boundary value on $\partial\mathrm{M}$, of
the term in the square brackets, vanishes.

We have therefore proved that $\left( -q^{2}\right)^{-\Delta}\mathcal{I}_{h,\bar{h}}\propto g$. Choosing a convenient normalization and going back to a general choice of $p$ we have arrived at the result (\ref{eq1001}).


\vfill

\eject

\bibliographystyle{plain}

\end{document}